\documentclass[lettersize,journal]{IEEEtran}
\usepackage{amsmath,amsfonts,amssymb}
\usepackage{dsfont}
\providecommand{\mathbbm}[1]{\mathds{#1}}
\usepackage{algorithmic}
\usepackage{algorithm}
\usepackage{array}
\usepackage[caption=false,font=normalsize,labelfont=sf,textfont=sf]{subfig}
\usepackage{textcomp}
\usepackage{stfloats}
\usepackage{url}
\usepackage{verbatim}
\usepackage{graphicx}
\usepackage{cite}
\usepackage{mdframed}
\usepackage{caption}
\usepackage{xcolor}

\makeatletter
\g@addto@macro\normalsize{%
  \setlength\abovedisplayskip{2pt}%
  \setlength\belowdisplayskip{2pt}%
  \setlength\abovedisplayshortskip{2pt}%
  \setlength\belowdisplayshortskip{2pt}%
}
\makeatother

\begin{document}

\title{RF‑GPT: Teaching AI to See the Wireless World}

\author{
Hang~Zou,
Yu~Tian,
Bohao~Wang,
Lina~Bariah,
Samson~Lasaulce,
Chongwen~Huang,
and~M\'erouane~Debbah

\thanks{H. Zou, Y. Tian, B. Wang,  L. Bariah and M. Debbah are with Research Institute for Digital Future, Khalifa University, 127788 Abu Dhabi, UAE (e-mails: \{hang.zou, yu.tian, lina.bariah, merouane.debbah\}@ku.ac.ae).
}
\thanks{B. Wang and C. Huang are with College of Information Science and Electronic Engineering, Zhejiang University, 310027, Hangzhou, China (email: \{bohaowang, chongwen.huang\}@zju.edu.cn)}
\thanks{S. Lasaulce is with Universit\'{e} de Lorraine, CNRS, CRAN, F-54000 Nancy, France (email: samson.lasaulce@univ-lorraine.fr).}}

\markboth{Journal of \LaTeX\ Class Files,~Vol.~14, No.~8, August~2021}%
{Shell \MakeLowercase{\textit{et al.}}: A Sample Article Using IEEEtran.cls for IEEE Journals}


\maketitle

\begin{abstract}
Large language models (LLMs) and multimodal models have become powerful general--purpose reasoning systems, yet radio–frequency (RF) signals, which underpin wireless systems, are still not natively supported. Existing LLM-based approaches for telecom focus mainly on text and structured data, while conventional RF deep-learning models are built separately for specific signal-processing tasks, highlighting a clear gap between RF perception and high-level reasoning. To bridge this gap, we introduce \textbf{RF-GPT}, a \emph{radio-frequency language model} (RFLM) that utilizes the visual encoders of multimodal LLMs to process and understand RF spectrograms. In this framework, complex in-phase/quadrature (IQ) waveforms are mapped to time–frequency spectrograms and then passed to pretrained visual encoders. The resulting representations are injected as RF tokens into a decoder-only LLM, which generates RF-grounded answers, explanations, and structured outputs. RF–GPT is trained via supervised instruction fine–tuning on a fully synthetic RF corpus, where standards–compliant waveform generators produce wideband scenes for six wireless technologies, from which we derive spectrograms, exact configuration metadata, and dense captions. A text–only LLM then converts these into RF–grounded instruction–answer pairs, yielding roughly 120{,}000 RF scenes and 0.625 million instruction examples without manual labeling. Across benchmarks for wideband modulation classification, overlap analysis, wireless–technology recognition, WLAN user counting, and 5G NR information extraction, RF–GPT achieves strong multi–task RF understanding, while general–purpose LLMs without RF grounding largely fail.
\end{abstract}



\section{Introduction}

\IEEEPARstart{L}{arge} language models (LLMs) have significantly advanced natural language processing, enabling powerful capabilities in long-form generation, code synthesis, tool use, and multi-step reasoning. Building on this, multimodal and vision--language models (VLMs) such as LLaVA~\cite{liu2023llava} and Qwen-VL~\cite{bai2025qwen25vl} integrate images and text for captioning, visual question answering, and grounded dialogue, while audio models like Whisper~\cite{radford2023whisper} show that large unlabeled speech/sound corpora can be leveraged for robust audio understanding. Despite these advances in text, vision, and audio, radio-frequency (RF) signals, which represent the physical layer for wireless communications, radar sensing, and integrated sensing-and-communications (ISAC), have not yet been integrated into these foundation model frameworks.

In parallel, the 6G research agenda envisions AI-native networks that tightly integrate sensing, communication, computing, and control, with autonomous, intent-driven operation across the radio access and core network~\cite{shahid2025large,6gIntelligentNetwork2024}. While existing machine learning-driven RF intelligence achieves high accuracy in specific tasks, it suffers from fragmentation, poor generalization, and a lack of interpretability, requiring distinct architectures and expensive datasets for each objective. Recent wireless foundation models such as WFM~\cite{cheraghinia2025-6g-wfm}, LWM~\cite{alikhani2024lwm}, WirelessGPT-like channel models~\cite{yang2025wirelessgpt}, and WiFo~\cite{liu2024wifo} move toward more generic RF representation learning, but in practice they still rely on task-specific heads or per-task fine-tuning to reach competitive performance on concrete downstream tasks. Each new application (e.g., channel estimation, localization, sensing, RF classification) typically requires a custom prediction head, loss, and fine-tuning pipeline, which undermines the promise of a truly universal RF foundation model. Moreover, many of these models are relatively small and optimized for a small set of benchmarks, making it difficult to balance performance across heterogeneous tasks. An LLM-based approach addresses these limitations by unifying multiple RF tasks into a single prompt-driven architecture. This shift enables shared representations across tasks and introduces a natural interface for reasoning, allowing models to justify predictions and interact with human operators.

Within this vision, LLMs have been explored as unified interfaces for knowledge access, reasoning, tool orchestration, and policy optimization, supporting automated fault diagnosis, configuration generation, and closed-loop network management. Early \emph{LLM4Telecom} efforts~\cite{bariah2024next,hao2025CST} and telecom-specific LLMs such as TelecomGPT~\cite{zou2025telecomgpt} embed domain knowledge to reason over alarms, KPIs, logs, and configuration tables. However, these systems remain largely \emph{text-centric}: they operate on tickets, logs, and structured KPIs, but do not directly ingest RF waveforms, spectrograms, or channel estimates. Beyond this text-only setting, recent wireless multi-task LLM frameworks such as LLM4WM~\cite{liu2025llm4wm}, WirelessAgent~\cite{tong2024wirelessagent}, the LLM-enabled multi-task physical-layer network of Zheng and Dai~\cite{zheng2026llmphy}, and the RF/CSI vision-language model LVM4CSI~\cite{guo2025lvm4csi} operate on wireless-domain inputs such as CSI, channel data, or image-like channel renderings, but expose downstream tasks through task-specific heads or per-task fine-tuning rather than producing free-form natural-language outputs. An emerging \emph{signal-token / LLM-processing} line further begins to bridge the modality gap by converting wireless signals into discrete token sequences ingested by a language model; however, it does not yet provide natural-language explanations or instruction-following on wideband RF scenes. This modality gap prevents LLM-style models from tackling RF-centric tasks that require direct access to the spectrum, such as identifying coexisting technologies, detecting harmful interference, analyzing occupancy patterns, or checking standard compliance at the signal level. Future AI-native networks will therefore require not only text-based reasoning but also deep, explainable RF perception.

These limitations motivate the development of \emph{radio-frequency language models} (RFLMs), which are foundation models that can process RF data and respond in natural language. In principle, an RFLM is a conditional language model grounded on RF tokens. Given an RF recording, it should be able to answer questions such as ``Which modulations and technologies are present in this band?'', ``Are there overlapping transmissions and what is their time–frequency relationship?'', or ``Is this behavior compliant with the relevant wireless standard?''. Beyond simple classification, an RFLM can provide textual explanations, structured JSON summaries for downstream controllers, and interactive dialogue with RF engineers or higher-level LLM agents. This language interface is aimed for practical reasons, including providing a unified instruction-following interface that comprises a wide range of RF downstream tasks, facilitating the integration of domain knowledge (e.g., ITU/3GPP standards, link-budget formulas, antenna constraints) and human feedback, as well as allowing human operators and autonomous agents to query, inspect, and manage RF behavior using plain language. In this sense, RF modeling becomes more flexible and interactive, rather than isolated predictors.

RF systems inherently generate large amounts of data (wideband monitoring, SDR deployments, cellular logs), and realistic simulators can produce vast amounts of labeled or semi-labeled signals under controlled conditions. This makes RF a promising candidate for self-supervised and synthetic pretraining. Using realistic waveform generators, we can generate signal-metadata samples, with accurate ground truth, such as modulation type, SNR, bandwidth, or resource allocation. Such a pipeline reduces the reliance on expert annotations and enables large-scale pretraining and instruction tuning. A pretrained RFLM can then be compressed or adapted for deployment across different RF scenarios, and can serve as an RF reasoning module across O-RAN, non-terrestrial networks, and ISAC scenarios, among other scenarios.

However, it is worth emphasizing that realizing such an RFLM is challenging. Unlike images and audio signals, RF waveforms are time-series, complex-valued signals sampled at very high rates, and understanding them depends on understanding time–frequency structure, protocol standards, and propagation effects. To the best of the authors' knowledge, large-scale datasets of real RF signals in diverse environment with expert textual annotations do not exist in the literature. Building such a dataset requires long-term RF measurements, specialized hardware, and domain experts for labeling. Over-the-air (OTA) data collection also raises privacy concerns, and typical RF environments are highly imbalanced, with rare but critical scenarios, such as extreme interference or unusual coexistence, being difficult to capture. 

In this paper, we propose \textbf{RF-GPT}, a radio-frequency language model that integrates RF spectrograms into a multimodal LLM, trained on a large synthetic RF instruction set with no human-annotated spectrogram labels and evaluated on a benchmark suite of five wideband RF understanding tasks. A comparison of RF-GPT against representative wireless foundation models, multi-task wireless LLM frameworks, and RF/CSI VLMs along the dimensions of input modality, backbone, training paradigm, head requirement, and output format is summarized in Table~\ref{tab:related_work}.

\begin{table*}[t!]
\centering
\caption{Positioning of RF-GPT against related works.}
\label{tab:related_work}
\renewcommand{\arraystretch}{1.0}
\setlength{\tabcolsep}{6pt}
\resizebox{\textwidth}{!}{%
\footnotesize
\begin{tabular}{p{2.5cm} p{2.0cm} p{2.8cm} p{2.4cm} c p{3.0cm}}
\hline
Model / family & Input modality & Backbone & Training paradigm & Heads? & Output format \\
\hline
WFM~\cite{cheraghinia2025-6g-wfm}            & IQ / channel    & Transformer encoder  & SSL + task heads  & yes & per-head logits \\
LWM~\cite{alikhani2024lwm}                   & CSI             & Transformer encoder  & SSL + task heads  & yes & per-head logits \\
WirelessGPT~\cite{yang2025wirelessgpt}       & CSI             & Transformer encoder  & Pretrain + FT     & yes & per-head logits \\
WiFo~\cite{liu2024wifo}                      & CSI             & Transformer encoder  & SSL (MAE)         & no  & numeric outputs \\
LLM4WM~\cite{liu2025llm4wm}                  & CSI             & LLM backbone         & Multi-task SFT    & yes & numeric outputs \\
WirelessAgent~\cite{tong2024wirelessagent}   & KPIs / network  & LLM backbone         & RAG / tool-use    & no  & text + actions \\
Zheng--Dai~\cite{zheng2026llmphy}            & PHY signals     & LLM backbone         & Multi-task FT     & yes & numeric outputs \\
LVM4CSI~\cite{guo2025lvm4csi}                & CSI (image)     & Pretrained LVMs      & Task-specific FT  & yes & numeric/label outputs \\
\textbf{RF-GPT (ours)}                       & Spectrogram     & ViT $+$ LLM          & RF-grounded SFT   & \textbf{no} & natural-language answer \\
\hline
\end{tabular}%
}
\end{table*}

The main contributions of this paper are summarized as follows:

\begin{itemize}
  \item We formulate the notion of a radio-frequency language model (RFLM) as a conditional language model grounded on RF tokens, and realize this concept through \textbf{RF-GPT}, which integrates RF spectrograms into a multimodal LLM via a vision encoder and a lightweight modality adapter through simple projection.
  \item We design standards‑compliant pipelines to synthesize a large RF spectrogram corpus for wideband settings and six wireless technologies, including 5G NR, LTE, UMTS, WLAN, DVB-S2, and Bluetooth, and introduce deterministic captioning schemes that convert signal‑level attributes into structured textual descriptions.
  \item We propose an automatic instruction-synthesis framework that converts RF captions into a diverse set of RF-grounded instruction–answer pairs, including explanation-style questions, quantitative queries (e.g., counts, overlaps), and structured JSON outputs, and we construct benchmarks that evaluate component recognition, overlap reasoning, technology identification, user counting, and NR-specific attribute extraction.
  \item We fine-tune pretrained multimodal LLMs on these synthetic RF instructions to obtain RF-GPT and demonstrate that RF grounding enables them to perform a range of RF understanding tasks on RF signals, including modulation classification, technology recognition, overlap analysis, and free-form RF question answering, while generic VLMs without RF grounding show very limited RF-specific understanding under our benchmark setting.
\end{itemize}

\section{Problem Formulation and RF-GPT Architecture}
\label{sec:problem_formulation}
In this section, we present the RF‑GPT architecture adopted in this work. Let $\boldsymbol{x} \in \mathbb{C}^{T}$ denote a complex baseband IQ sequence of length $T$ samples, and let 
$\boldsymbol{y} = (y_{1},\dots,y_{N})$ denote a text sequence of $N$ tokens from a vocabulary $\mathcal{V}$. This text sequence is generally relevant to the RF signals, e.g., a description of their characteristics or intents associated with them. We denote by $\phi_{\mathrm{RF}}$ an RF encoder that maps the raw IQ sequence into a sequence of 
$M$ RF tokens in a latent space, 
\begin{equation}
\phi_{\mathrm{RF}} : \mathbb{C}^{T} \rightarrow \mathbb{R}^{M \times d},
\end{equation}
where $d$ is the embedding dimension of each RF token. In our formulation, an RFLM is a conditional language model that generates a token sequence $\boldsymbol{y}$ given an RF input $\boldsymbol{x}$ that has been encoded into a sequence of RF tokens $\phi_{\mathrm{RF}}(\boldsymbol{x})$. The model defines the autoregressive conditional distribution as
\begin{equation}
\mathcal{P}_{\Theta}(\boldsymbol{y}\mid \phi_{\mathrm{RF}}(\boldsymbol{x}))
=\prod_{t=1}^{N}
\mathcal{P}_{\Theta}\bigl(y_{t}\mid \boldsymbol{y}_{<t}, \phi_{\mathrm{RF}}(\boldsymbol{x})\bigr),
\label{eq:rflm}
\end{equation}
where $\boldsymbol{y}_{<t} = (y_1,\dots,y_{t-1})$ and $\Theta$ collects all trainable parameters of RF-GPT. The RF tokens serve as a prefix that conditions an autoregressive language model on the RF input.  

\begin{figure}[t!]
\centering
\includegraphics[width=1.0\linewidth]{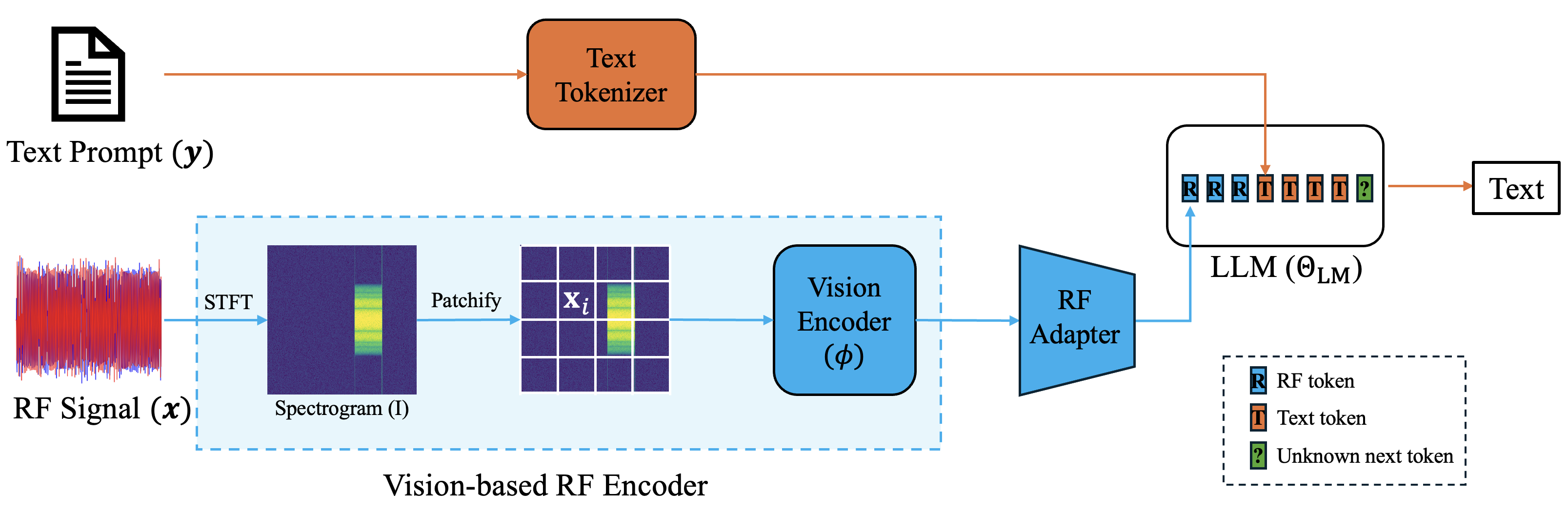}
\vspace{-4mm}
\caption{Basic structure of RF-GPT.}
\label{fig:rf_gpt_structure}
\end{figure}

The basic structure of RF-GPT is illustrated in Fig.~\ref{fig:rf_gpt_structure}. RF-GPT consists of (i) an RF encoder, (ii) an MLP adapter that projects RF tokens into the language-model embedding space, and (iii) a decoder-only LLM that produces RF-grounded text outputs. In the following, we describe each component and the RF-grounded supervised fine-tuning (SFT) procedure.

\subsection{RF Encoding via Spectrograms and Vision Encoder}
\label{subsec:rf_encoder}
Applying Transformers directly to raw IQ sequences sampled at RF rates is computationally infeasible \cite{vaswani2017attention} and ignores the strong inductive bias of time--frequency structure \cite{tian2026survey}. Hence, we first map an RF signal $\boldsymbol{x}$ to a two-dimensional time--frequency representation using a short-time Fourier transform (STFT). Let the sampling rate be denoted by $f_s$, and let $x[n]$ represent the $n$-th complex sample at time $t = n/f_s$. We compute the STFT with window $w[n]$, fast Fourier transform (FFT) size $F$, and hop size $H_\text{hop}$,
\begin{equation}
S[k,t] \;=\; \sum_{n} x[n]\, w[n-tH_\text{hop}]\, e^{-j\frac{2\pi}{F}kn}, k\in[0,F{-}1],
\end{equation}
where $t$ represents the time index and $S[k,t] \in \mathbb{C}$ is the complex STFT coefficient at frequency bin $k$ and frame $t$. We then convert the complex spectrogram to a magnitude (or power) spectrogram in dB by applying log compression, as follows
\begin{equation}
    A_\text{dB}[k,t] = 10\log_{10} \bigl(|S[k,t]|^2+\varepsilon\bigr),
\end{equation}
where $\varepsilon > 0$ is a small constant for numerical stability. The resulting time--frequency matrix is normalized and mapped to an image $\mathrm{I} \in \mathbb{R}^{H \times W \times C}$, where $H$ and $W$ are the image height and width and $C$ is the number of channels. We either consider $A_\text{dB}$ as a single-channel grayscale image ($C=1$) or apply a fixed colormap (e.g., viridis) to obtain a pseudo-RGB image ($C=3$). This yields an RF spectrogram image that can be processed by a standard vision encoder $\mathcal{E}_{\mathrm{v}}$. We define the complete RF encoder $\phi_{\mathrm{RF}}$ as the composition of the spectrogram pipeline and the vision pipeline, as
\begin{equation}\phi_{\mathrm{RF}}(\boldsymbol{x}) = \mathcal{E}_{\mathrm{v}}(\mathrm{I})=\mathcal{E}_{\mathrm{v}}(\mathrm{Spec}(\boldsymbol{x})),\end{equation}where $\mathrm{Spec}(\cdot)$ denotes the spectrogram pipeline (STFT + magnitude/dB + colormap).

\textbf{Patch embedding.}
The spectrogram image \(\mathrm{I}\) is partitioned into a regular grid of non-overlapping patches of size
$P \times P$ pixels, where both $H$ and $W$ are divisible by $P$. The number of patches (and thus RF tokens) is \(M = \frac{H}{P} \cdot \frac{W}{P}\).
Let $\mathbf{x}_i \in \mathbb{R}^{P^2 C}$ denote the vector obtained by flattening the
$i$-th patch (row-major order), for $i = 1,\dots,M$. Each patch is linearly projected to a $d$-dimensional embedding
using a learnable matrix $\mathbf{E} \in \mathbb{R}^{d \times (P^2 C)}$,
\begin{equation}
\mathbf{e}_i = \mathbf{E} \mathbf{x}_i \in \mathbb{R}^{d}, \qquad i = 1,\dots,M.
\end{equation}
This patchification and linear projection step acts as an RF tokenizer, turning each spectrogram into a sequence of \(M\) RF tokens that the language model can consume.
We also use a learnable positional embedding $\mathbf{p}_i \in \mathbb{R}^{d}$
for each patch index $i$, which encodes its 2D location on the spectrogram
grid (time--frequency position). The input token sequence to the transformer encoder is then given by
\begin{equation}
\mathbf{z}_i^{(0)} = \mathbf{e}_i + \mathbf{p}_i, \qquad i = 1,\dots,M,
\end{equation}
and we stack them into
\begin{equation}
\mathbf{Z}^{(0)} =
\begin{bmatrix}
(\mathbf{z}_1^{(0)})^\top \\
\vdots \\
(\mathbf{z}_M^{(0)})^\top
\end{bmatrix}
\in \mathbb{R}^{M \times d}.
\end{equation}
Unlike some ViT variants \cite{dosovitskiy2020image,touvron2021training}, we do not prepend a special classification token, instead, all $M$ patch tokens are treated as RF tokens and later passed to the language model.

\textbf{Transformer encoder layers.}
The matrix $\mathbf{Z}^{(0)}$ is processed by $L$ stacked transformer encoder blocks. Each block consists of multi-head self-attention (MHA) followed by a position-wise multi-layer perceptron (MLP), both wrapped with residual connections and layer normalization. Given $\mathbf{Z}^{(\ell)} \in \mathbb{R}^{M \times d}$ as the input to
layer $\ell$ ($\ell = 0,\dots,L-1$), the output $\mathbf{Z}^{(\ell+1)}$ is obtained as
\begin{align}
\widetilde{\mathbf{Z}}^{(\ell)} 
&= \mathbf{Z}^{(\ell)} 
   + \mathrm{MHA}\bigl(\mathrm{LN}(\mathbf{Z}^{(\ell)})\bigr), \label{eq:vit_block_attn} \\
\mathbf{Z}^{(\ell+1)} 
&= \widetilde{\mathbf{Z}}^{(\ell)} 
   + \mathrm{MLP}\bigl(\mathrm{LN}(\widetilde{\mathbf{Z}}^{(\ell)})\bigr),
   \label{eq:vit_block_mlp}
\end{align}
where $\mathrm{LN}(\cdot)$ is layer normalization and the MLP is a
two-layer feed-forward network (FFN) with nonlinearity, applied row-wise as
\begin{equation}
\mathrm{MLP}(\mathbf{z}) = \sigma\bigl(\mathbf{z}\mathbf{W}_1 + \mathbf{b}_1\bigr) \mathbf{W}_2 + \mathbf{b}_2,
\end{equation}
for learnable weights $\mathbf{W}_1, \mathbf{W}_2$ and biases $\mathbf{b}_1, \mathbf{b}_2$, and activation function $\sigma(\cdot)$ (e.g., GELU). 

The multi-head self-attention (MHA) operates as follows. For a given input $\mathbf{U} \in \mathbb{R}^{M \times d}$, the $h$-th attention head computes
\begin{align}
\mathbf{Q}_h &= \mathbf{U} \mathbf{W}_h^{Q}, \quad
\mathbf{K}_h = \mathbf{U} \mathbf{W}_h^{K}, \quad
\mathbf{V}_h = \mathbf{U} \mathbf{W}_h^{V}, \\
\mathbf{A}_h &= \mathrm{softmax}\!\left(\frac{\mathbf{Q}_h \mathbf{K}_h^\top}{\sqrt{d_k}}\right),  \quad
\mathbf{H}_h = \mathbf{A}_h \mathbf{V}_h,
\end{align}
where $\mathbf{W}_h^{Q}, \mathbf{W}_h^{K}, \mathbf{W}_h^{V} \in \mathbb{R}^{d \times d_k}$
are learnable projection matrices (query, key, and value for head $h$), $d_k$ is the head dimension, and $\mathrm{softmax}$ is applied row-wise over the attention scores. The outputs of all heads are concatenated and linearly projected, as follows
\begin{equation}
\mathrm{MHA}(\mathbf{U}) = 
\bigl[\mathbf{H}_1 \,\|\, \mathbf{H}_2 \,\|\, \dots \,\|\, \mathbf{H}_H\bigr] \mathbf{W}^{O},
\end{equation}
where $\mathbf{W}^{O} \in \mathbb{R}^{(H d_k) \times d}$ is the projection matrix and $H$ represents the number of attention heads. After $L$ layers, the final patch embeddings are represented as
\begin{equation}
\boldsymbol{h}_{i} = \mathbf{z}_i^{(L)} \in \mathbb{R}^{d}, 
\qquad i = 1,\dots,M.
\end{equation}
These vectors form the RF-grounded latent sequence $\boldsymbol{h}_{1:M}$ used for downstream conditioning. 

\textbf{Remark.}
In our framework, the vision encoder \emph{serves as the RF encoder} by treating the spectrogram as an image and encoding time--frequency patterns such as modulation structure, resource allocation, interference, and coexistence into a sequence of latent RF tokens. There are two main motivations for transforming a vision encoder into an RF encoder, namely, (i) RF signals span very different carrier frequencies, bandwidths, and sampling rates across technologies, making it difficult to design a single raw-IQ tokenizer that is simultaneously efficient and robust for narrowband, wideband, and multi-standard scenarios, and (ii) the STFT serves as an effective RF feature extractor, while a vision encoder pretrained on large-scale image datasets can extract universal semantic structure from images, including spectrograms.

A key design decision in RF-GPT is to operate on magnitude spectrograms rather than raw IQ signals. From a time--frequency analysis viewpoint, the spectrogram is a smoothed energy distribution associated with the signal, and can be seen as a windowed version of Wigner--Ville type representations \cite{flandrin1999time}. As such, many RF attributes of interest, e.g., occupied bandwidth, center frequency and Doppler shifts, burst timing, time--frequency sparsity patterns, and even modulation structure, are encoded geometrically in the spectrogram and are therefore accessible to a vision encoder.

It is important to distinguish between two regimes. First, \emph{true phase retrieval} aims to reconstruct the entire complex baseband waveform from magnitude-only measurements (e.g., STFT magnitude). This problem is nontrivial, but the phase-retrieval literature shows that, under suitable redundancy and support assumptions, a signal can be uniquely determined (up to a global phase) and stably recovered from such magnitude data; see, e.g., work on STFT phase retrieval \cite{jaganathan2016stft} and convex formulations such as PhaseLift \cite{candes2013phaselift}. Second, in our setting we do not attempt to reconstruct $x[n]$ itself, but only to infer task-relevant attributes such as modulation family, technology, SNR regime, or overlap structure. These tasks are many-to-one mappings of the waveform and depend primarily on the time--frequency energy patterns, not on the exact sample-wise phase. RF-GPT therefore uses magnitude-only spectrograms as a rich but lossy front-end representation. While they are not sufficient for all possible RF tasks, e.g., those requiring absolute phase information or precise multi-antenna phase relationships, they retain enough structure to support the perception-oriented tasks considered in this work. Beyond this trade-off, three properties make the spectrogram well-suited here: it is universal across any standard and sample rate, efficient to compute as a deterministic STFT with no learned IQ or CSI encoder, and jointly time--frequency, exposing burst timing, bandwidth, modulation envelope, and coexistence to a pretrained vision backbone. In Sec.~\ref{sec:method} we empirically validate that this spectrogram-based approach allows the model to recover the RF attributes required by our benchmarks with high accuracy. With the RF encoding pipeline established, we now describe how the resulting RF tokens are used to condition the language model and ground its predictions in RF information.

\subsection{Language Model Architecture and RF Conditioning}
\label{subsec:lm_backbone}

The backbone of RF-GPT is a decoder-only Transformer language model with parameters $\Theta_{\mathrm{LM}}$, which are part of the overall parameter set $\Theta$. The LLM operates on a sequence of token embeddings
$\mathbf{u}_{1:(M+N)} \in \mathbb{R}^{(M+N)\times d_{\mathrm{LM}}}$,
where $d_{\mathrm{LM}}$ is the model dimension of the LLM. We add standard positional embeddings e.g., rotary position embeddings (RoPE \cite{su2024rope}) to $\mathbf{u}_{1:(M+N)}$ before passing them to the decoder layers.

First, the RF encoder produces $M$ visual tokens
$\boldsymbol{h}_{1:M} = \mathcal{E}_{\mathrm{v}}(I)$, with
$\boldsymbol{h}_{i} \in \mathbb{R}^{d}$ as defined in the
previous subsection. An RF adapter through simple projection maps these to
the LLM dimension:
\begin{equation}
{
\mathbf{r}_{i}
= \mathbf{W}_{2}\, \sigma\!\left(\mathbf{W}_{1} \boldsymbol{h}_{i} + \mathbf{b}_{1}\right) + \mathbf{b}_{2}
\in \mathbb{R}^{d_{\mathrm{LM}}}, \quad i = 1,\dots,M,
}
\end{equation}
where $\mathbf{W}_{1} \in \mathbb{R}^{d \times d}$, $\mathbf{b}_{1} \in \mathbb{R}^{d}$, $\mathbf{W}_{2} \in \mathbb{R}^{d_{\mathrm{LM}} \times d}$, and $\mathbf{b}_{2} \in \mathbb{R}^{d_{\mathrm{LM}}}$ are trainable parameters, and $\sigma(\cdot)$ is a non-linearity (e.g., GELU). This adapter plays the role of a visual–language adapter, similar to the projection modules used in LLaVA-style models \cite{liu2023llava}, mapping the vision-encoder features (which encode RF information) into the LLM embedding space.

Text tokens $y_t$ are embedded via a standard lookup table
$\mathbf{E}_{\mathrm{tok}} \in \mathbb{R}^{|\mathcal{V}| \times d_{\mathrm{LM}}}$:
\begin{equation}
\mathbf{e}_{t} = \mathbf{E}_{\mathrm{tok}}[y_t] \in \mathbb{R}^{d_{\mathrm{LM}}}, \quad t = 1,\dots,N,
\end{equation}
with $|\mathcal{V}|$ the vocabulary size of the LLM backbone. We then form a single joint sequence by
concatenating RF and text embeddings:
\begin{equation}
\mathbf{u}_{1:(M+N)} =
\bigl[\,\mathbf{r}_{1},\dots,\mathbf{r}_{M},
       \mathbf{e}_{1},\dots,\mathbf{e}_{N}\,\bigr].
\end{equation}

Let $\mathbf{U}^{(0)} \in \mathbb{R}^{(M+N)\times d_{\mathrm{LM}}}$ be the
matrix whose rows are $\mathbf{u}_{1}, \dots,\mathbf{u}_{M+N}$. The sequence is processed by
$L_{\mathrm{LM}}$ stacked decoder layers. Each layer follows a modern
\emph{pre-normalized} Transformer design with root mean square (RMS) normalization, grouped-query
attention (GQA) with a causal mask, and a gated Up/Down MLP, as adopted in
recent Qwen-series LLMs. For layer $\ell = 0,\dots,L_{\mathrm{LM}}-1$ we write
\begin{align}
\widetilde{\mathbf{U}}^{(\ell)}
&= \mathbf{U}^{(\ell)}
   + \mathrm{GQA}_{\mathrm{causal}}\!\bigl(
        \mathrm{RMSNorm}\bigl(\mathbf{U}^{(\ell)}\bigr)
     \bigr),
   \label{eq:lm_block_attn} \\
\mathbf{U}^{(\ell+1)}
&= \widetilde{\mathbf{U}}^{(\ell)}
   + \mathrm{MLP}_{\mathrm{gated}}\!\bigl(
        \mathrm{RMSNorm}\bigl(\widetilde{\mathbf{U}}^{(\ell)}\bigr)
     \bigr),
   \label{eq:lm_block_mlp}
\end{align}
where $\mathrm{RMSNorm}(\cdot)$ is root-mean-square layer normalization \cite{zhang2019root},
$\mathrm{GQA}_{\mathrm{causal}}$ is grouped-query self-attention \cite{ainslie2023gqa} with a
causal mask, and $\mathrm{MLP}_{\mathrm{gated}}$ is a gated feed-forward block
(Up/Down MLP) \cite{shazeer2020glu}. Since decoder-only transformers and these building blocks are not yet standard in RF modeling, we briefly recall their definitions.

\textbf{RMS normalization.}
For a token vector $\mathbf{x} \in \mathbb{R}^{d_{\mathrm{LM}}}$, RMSNorm is
defined as
\begin{equation}
\mathrm{RMSNorm}(\mathbf{x})
= \gamma \odot \frac{\mathbf{x}}{\sqrt{\tfrac{1}{d_{\mathrm{LM}}}
  \lVert \mathbf{x} \rVert_2^2 + \varepsilon}},
\end{equation}
where $\gamma \in \mathbb{R}^{d_{\mathrm{LM}}}$ is a learned scale,
$\varepsilon>0$ is a small constant, and $\odot$ denotes element-wise
multiplication. RMSNorm stabilizes activations while being slightly simpler
than standard LayerNorm.

\textbf{Grouped-query Attention (GQA).}
Let $\mathbf{U} \in \mathbb{R}^{(M+N) \times d_{\mathrm{LM}}}$ denote a generic layer input. Grouped-query attention uses $H_q$ query heads and
$H_k$ key–value heads with $H_q \ge H_k$. We first compute
\begin{align}
\mathbf{Q} &= \mathbf{U} \mathbf{W}^{Q}, \quad
\mathbf{K} = \mathbf{U} \mathbf{W}^{K}, \quad
\mathbf{V} = \mathbf{U} \mathbf{W}^{V},
\end{align}
where
$\mathbf{W}^{Q} \in \mathbb{R}^{d_{\mathrm{LM}} \times H_q d_k}$ and
$\mathbf{W}^{K}, \mathbf{W}^{V} \in \mathbb{R}^{d_{\mathrm{LM}} \times (H_k d_k)}$ are
learnable projection matrices, and $d_k$ is the head dimension.
To facilitate grouped-query computation, we reshape $\mathbf{Q}$ into $H_q$ separate heads and $\mathbf{K}, \mathbf{V}$ into $H_k$ groups. Formally, let $\mathbf{Q}_{h} \in \mathbb{R}^{(M+N) \times d_{k}}$ denote the $h$-th query head for $h \in \{1, \dots, H_q\}$, and let $\mathbf{K}_{g(h)}, \mathbf{V}_{g(h)} \in \mathbb{R}^{(M+N) \times d_{k}}$ represent the key and value heads shared by the group associated with query head $h$.
The grouping is determined by the index function $g(h) = \lfloor(h-1)/r\rfloor + 1$, where $r = H_q / H_k$ is the group size, mapping the $h$-th query head to its corresponding key-value group index. With a causal attention mask
$M_{\mathrm{causal}} \in \mathbb{R}^{(M+N) \times (M+N)}$ (entries $0$ for allowed
positions and $-\infty$ otherwise), GQA is
\begin{align}
\mathbf{A}_h &= \mathrm{softmax}\!\left(
  \frac{\mathbf{Q}_h \mathbf{K}_{g(h)}^\top}{\sqrt{d_k}} + M_{\mathrm{causal}}
\right), \nonumber
\\ 
\mathbf{H}_h &= \mathbf{A}_h \mathbf{V}_{g(h)},
\end{align}
and the heads are concatenated and linearly projected back to
$d_{\mathrm{LM}}$:
\begin{equation}
\mathrm{GQA}_{\mathrm{causal}}( \mathbf{U})
= \bigl[\,\mathbf{H}_1 \,\|\, \dots \,\|\, \mathbf{H}_{H_q}\,\bigr] \mathbf{W}^{O},
\end{equation}
with $\mathbf{W}^{O} \in \mathbb{R}^{(H_q d_k) \times d_{\mathrm{LM}}}$.
The causal mask enforces that position $t$ attends only to positions
$\leq t$, ensuring autoregressive generation.

\textbf{Gated Up/Down MLP.}
The feed-forward block uses a gated Up/Down structure (often instantiated
with a SwiGLU-style activation) instead of a plain two-layer MLP. For an
input matrix $\mathbf{X} \in \mathbb{R}^{T \times d_{\mathrm{LM}}}$, we compute
\begin{align}
\mathbf{U}_{\mathrm{up}}   &= \mathbf{X} \mathbf{W}_{\mathrm{up}}   + \mathbf{b}_{\mathrm{up}},
\\
\mathbf{U}_{\mathrm{gate}} &= \mathbf{X} \mathbf{W}_{\mathrm{gate}} + \mathbf{b}_{\mathrm{gate}},
\end{align}
where $\mathbf{W}_{\mathrm{up}}, \mathbf{W}_{\mathrm{gate}} \in
\mathbb{R}^{d_{\mathrm{LM}} \times d_{\mathrm{ff}}}$ project into a larger
hidden dimension $d_{\mathrm{ff}}$, and
$\mathbf{b}_{\mathrm{up}}, \mathbf{b}_{\mathrm{gate}} \in \mathbb{R}^{d_{\mathrm{ff}}}$.
A pointwise nonlinear gating (e.g., SwiGLU) is then applied:
\begin{align}
\mathbf{G} &= \phi_\mathrm{SwiGLU}\left(\mathbf{U}_{\mathrm{gate}}\right) \odot \mathbf{U}_{\mathrm{up}}, 
\\
\mathrm{MLP}_{\mathrm{gated}}(\mathbf{X})
&= \mathbf{G} \mathbf{W}_{\mathrm{down}} + \mathbf{b}_{\mathrm{down}},
\end{align}
with $\mathbf{W}_{\mathrm{down}} \in \mathbb{R}^{d_{\mathrm{ff}} \times d_{\mathrm{LM}}}$
and $\mathbf{b}_{\mathrm{down}} \in \mathbb{R}^{d_{\mathrm{LM}}}$. The gating
mechanism improves expressivity at similar or lower computational cost than a
standard two-layer MLP.

Finally, the last decoder layer produces
$\mathbf{U}^{L_{\mathrm{LM}}} \in \mathbb{R}^{(M+N)\times d_{\mathrm{LM}}}$.
We take the final $N$ positions corresponding to the text tokens (indices $M+1$ to $M+N$) and apply a linear output head and softmax to obtain the conditional distribution
$\mathcal{P}_{\Theta}(\boldsymbol{y}\mid \phi_{\mathrm{RF}}(\boldsymbol{x}))$ in
\eqref{eq:rflm}. Apart from the RF tokens being concatenated as a prefix,
the LLM follows a modern architecture of LLMs with grouped-query attention,
RMSNorm, and gated MLPs (e.g., Qwen- and LLaMA-style models).

\subsection{RF-Grounded Supervised Fine-Tuning}
\label{subsec:rf_grounded_sft}

With the model structure of RF-GPT in place, we now discuss how to inject knowledge of RF signals into it. To adapt a generic VLM into an RFLM, we perform RF-grounded supervised fine-tuning (SFT) on a synthetic instruction dataset built from standards-compliant waveform generators. We assume access to a dataset \(\mathcal{D}_{\mathrm{RF}} = \bigl\{(\boldsymbol{x}^{(i)}, \boldsymbol{q}^{(i)}, \boldsymbol{y}^{(i)})\bigr\}_{i=1}^{K}\),
where $\boldsymbol{x}^{(i)}$ is an RF waveform (IQ samples) that can be converted into a spectrogram image as described in Sec.~\ref{subsec:rf_encoder},
$\boldsymbol{q}^{(i)}$ is a natural-language instruction or question about the corresponding spectrogram
(e.g., ``Describe the signal types and overlaps in this RF scene.''), and $\boldsymbol{y}^{(i)}$ is the desired answer (e.g., a caption, explanation, or JSON summary). These triplets are constructed in two stages, namely, RF spectrogram captioning and RF instruction synthesis (see Sec.~\ref{sec:rf_grounding}). Given this dataset, RF-GPT is trained to minimize the standard autoregressive cross-entropy loss over the answer tokens, conditioned on both the RF tokens and the instruction, as follows
\begin{equation}
\mathcal{L}(\Theta)
= - \sum_{(\boldsymbol{x}, \boldsymbol{q}, \boldsymbol{y}) \in \mathcal{D}_{\mathrm{RF}}}
   \sum_{t=1}^{|\boldsymbol{y}|} 
   \log \mathcal{P}_{\Theta}\Bigl(
     y_{t} \,\big|\, \boldsymbol{y}_{<t}, \boldsymbol{q}, \phi_{\mathrm{RF}}(\boldsymbol{x})
   \Bigr).
\label{eq:sft_loss}
\end{equation}
In practice, we use the usual instruction-tuning format in which $(\boldsymbol{q}, \boldsymbol{y})$ are 
concatenated into a single text sequence with appropriate role markers, the RF tokens $\phi_{\mathrm{RF}}(\boldsymbol{x})$ are injected as a prefix, and gradients are applied only on the answer tokens $\boldsymbol{y}$. The key point is that \emph{all} supervision is RF-grounded, by construction, every target token in $\boldsymbol{y}$ must be consistent with the underlying RF scene, $\boldsymbol{x}$. This encourages the model to align RF patterns in the spectrogram with RF concepts in language compared to generic VLMs, which exhibit very limited RF-specific understanding under our benchmark setting, as we will demonstrate in Sec.~\ref{sec:method}.

\section{RF Grounding and Benchmarking}
\label{sec:rf_grounding}

Training an RFLM requires a high-quality RF spectrogram-caption dataset, which is difficult to assemble from real measurements. Accurate captions demand domain expertise, manual labeling is expensive and unreliable, and the ground truth is often hidden inside proprietary base-station schedulers and inaccessible to an external receiver. Real campaigns also need specialized hardware and planning, and the data they yield is highly imbalanced, with rare but critical edge cases barely observed. Achieving both label quality and statistical coverage from real data would require an industrial-scale effort beyond the scope of typical academic work. We therefore use synthetic data from standards-compliant generators, which provide exact ground truth for every latent variable of interest, such as modulation, bandwidth, and resource allocation. They let us sweep technologies, numerologies, traffic patterns, coexistence, and impairment levels to reach a diversity and coverage real data cannot cheaply match, and they yield precise, unambiguous captions of the full configuration, exactly the supervision an RFLM needs.  

\subsection{Wireless Technology Waveform Generation}
\label{subsec:waveform_generation}

We use MATLAB's Wireless Waveform Generator family as a unified tool to generate synthetic, realistic RF waveforms. Rather than describing each technology pipeline independently, we follow a common workflow that we design for six major wireless technologies, namely, 5G NR, LTE, UMTS, WLAN, DVB-S2, and Bluetooth. 

\begin{itemize}
  \item \textbf{Technology and configuration sampling:} We first select a target technology (e.g., 5G NR) and create a base configuration object using the corresponding MATLAB API. From this base, we define a bounded but rich parameter space (bandwidth, numerology, etc.) and randomly sample candidate configurations.
  \item \textbf{Standards-compliant waveform generation:} Each sampled configuration is passed to the toolbox waveform generator, which implements the relevant 3GPP/IEEE physical-layer specifications. Configurations that violate standard constraints are rejected and resampled, ensuring that all accepted waveforms are standards compliant.
  \item \textbf{Channel and impairment modeling (optional):} We optionally apply channel models and RF impairments and adjust noise levels to target specific SNR ranges, recording all parameters as part of the metadata.
  \item \textbf{Spectrogram construction:} The resulting complex baseband IQ waveform is converted to a time--frequency representation via STFT, mapped to a fixed-resolution grayscale or pseudo-RGB spectrogram image, and used as input to the RF encoder.
  \item \textbf{Metadata logging:} For every sample, we store the spectrogram image, the full MATLAB configuration object, the channel/noise parameters, and derived attributes, e.g., link direction and number of users, providing machine-readable ground truth for captioning, instruction synthesis, and benchmarking.
\end{itemize}

\begin{figure}[t!]
\centering
\includegraphics[width=0.85\linewidth]{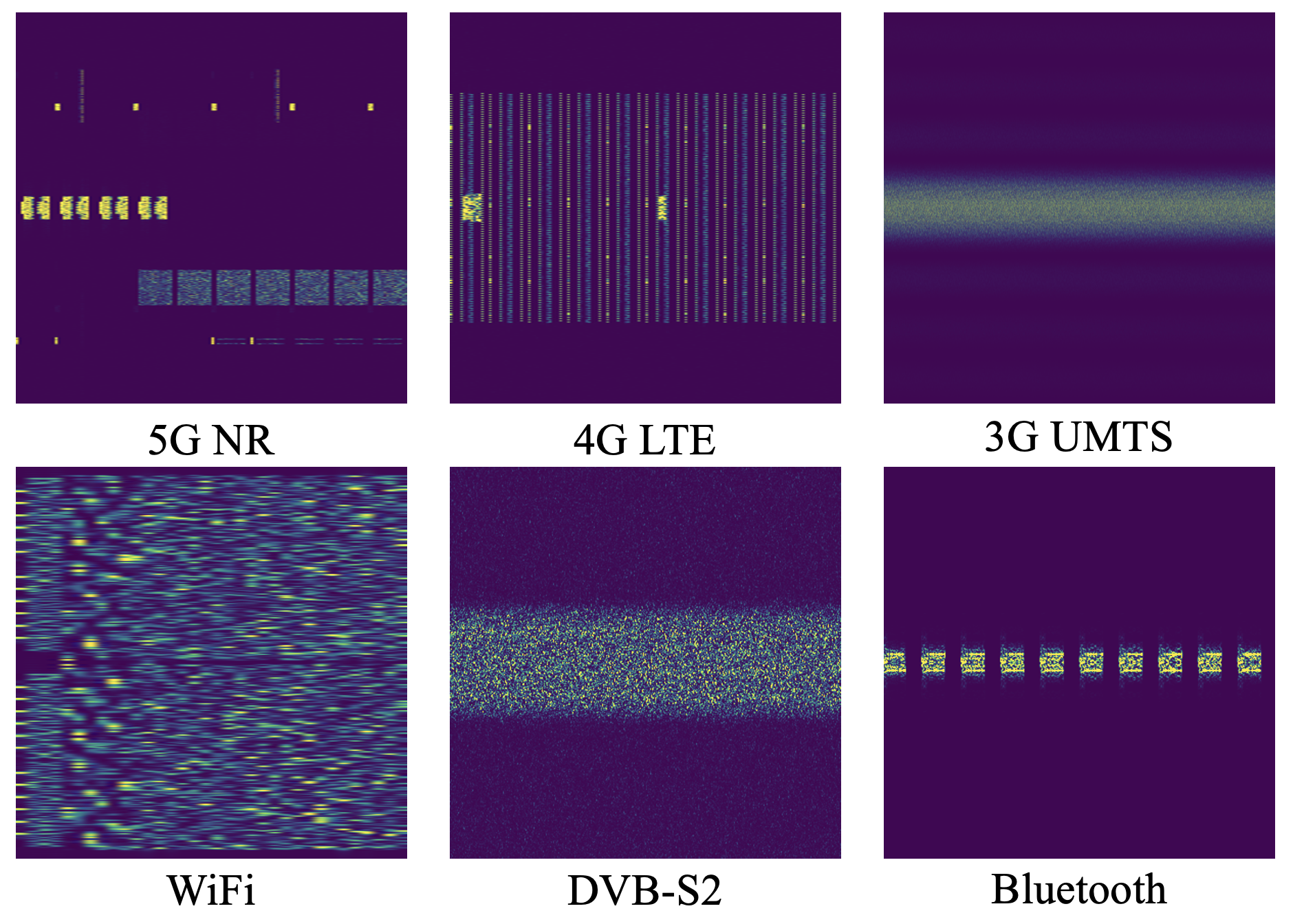}
\vspace{-2mm}
\caption{Wireless technologies gallery: time-frequency spectrograms from six important RF signals including 5G NR, 4G LTE, 3G UMTS, WLAN, DVB-S2 and Bluetooth.}
\label{fig:wireless_tech_gallery}
\end{figure}

This common workflow is designed for each of the six considered technologies, with technology-specific APIs, parameter ranges, and validation checks. Together, these six technology-specific generators give us a unified, standards-compliant synthetic datasets that covers a broad range of the wireless scenarios. Because every spectrogram is generated from a known configuration object and we keep that object verbatim, we obtain exact and noise-free labels for modulation type, bandwidth, numerology, scheduling, and link-level structure annotations. Examples of each wireless technology are provided in Fig. \ref{fig:wireless_tech_gallery} for reference. These standardized waveforms and their metadata form the basis for the captioning and instruction‑synthesis pipeline described next.

\begin{figure}[t!]
\centering
\includegraphics[width=1.0\linewidth]{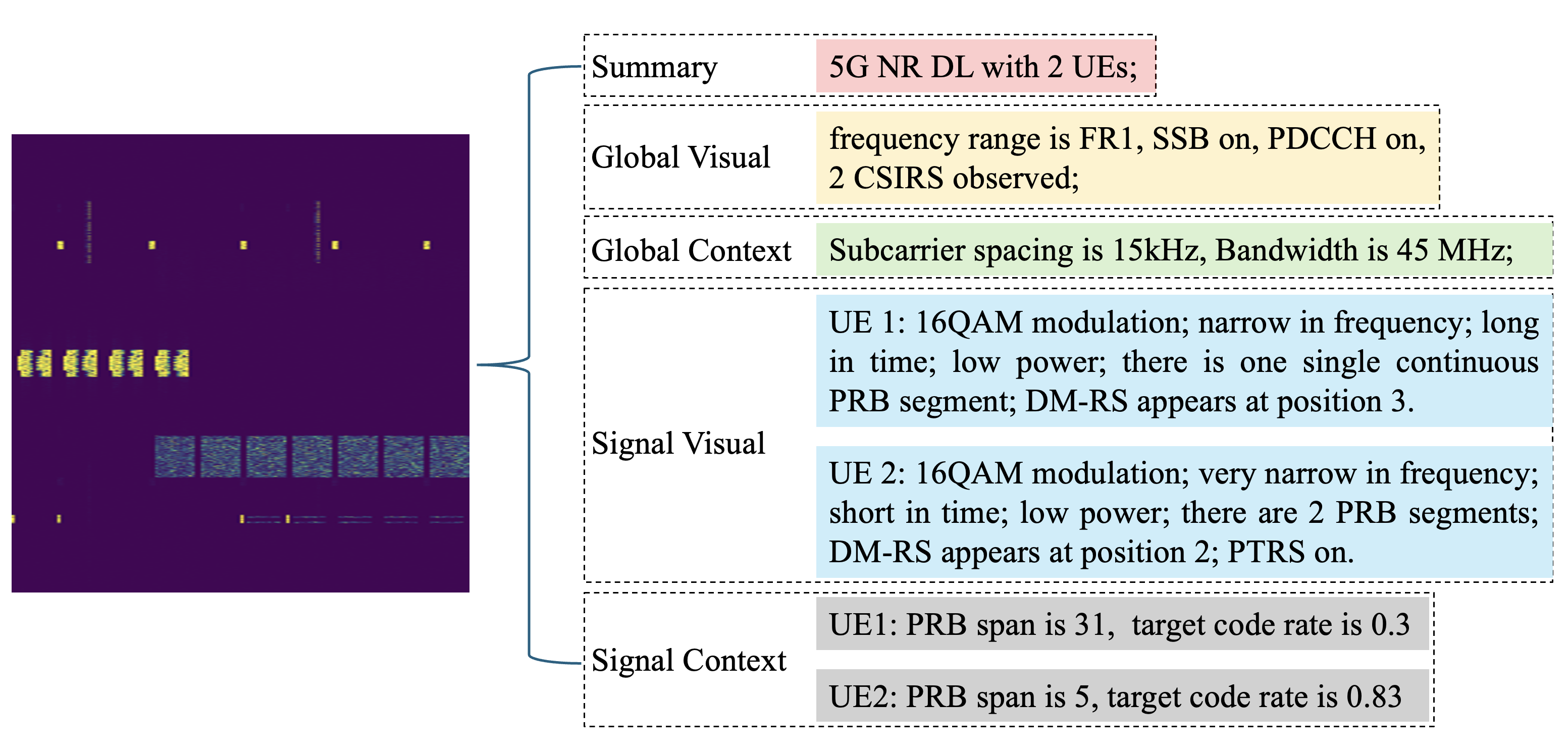}
\vspace{-4mm}
\caption{Our fine-grained caption strategy for RF spectrograms.}
\label{fig:fine_grained_caption_strategy}
\end{figure}

\begin{figure}[t!]
\centering
\includegraphics[width=1.0\linewidth]{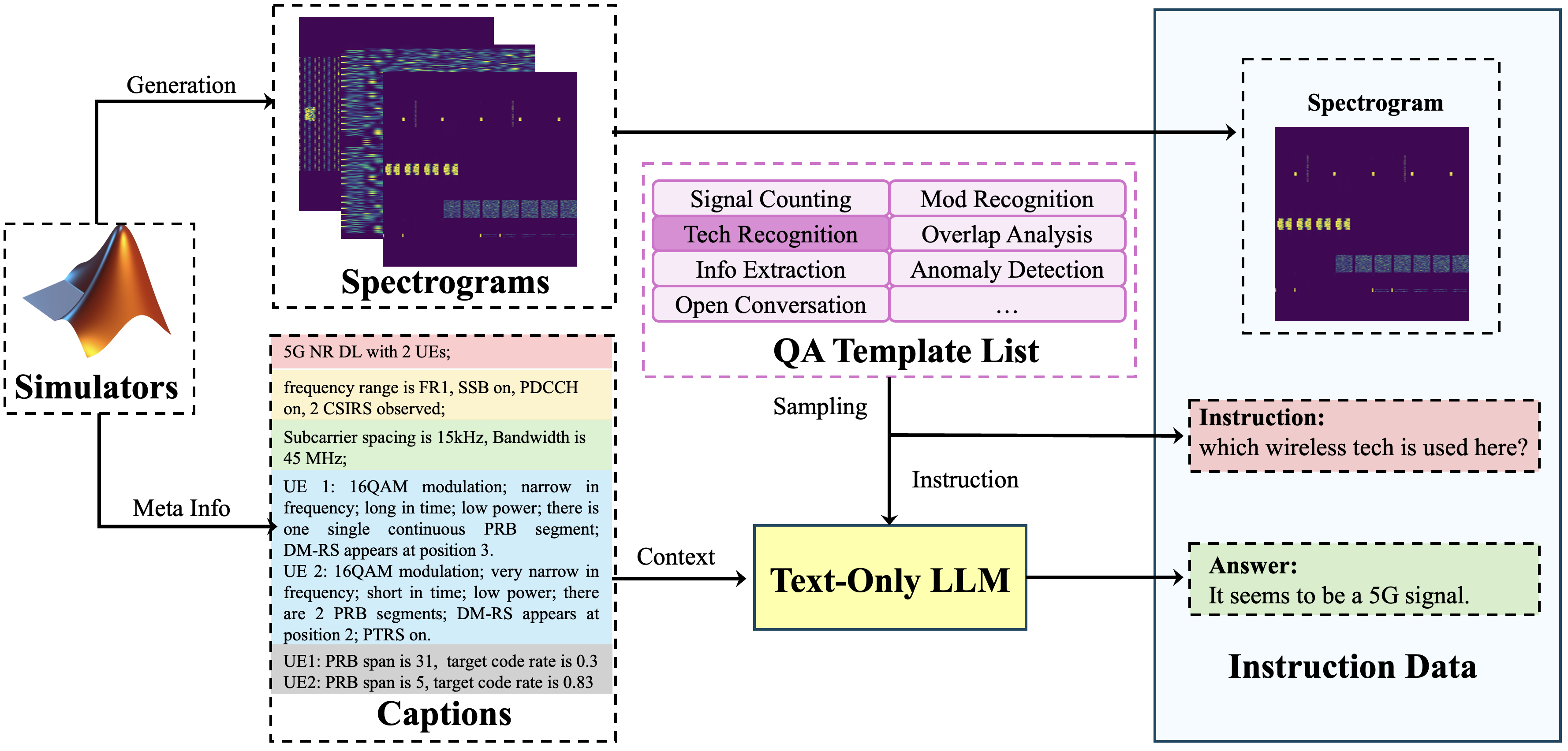}
\vspace{-4mm}
\caption{RF-grounded instruction synthesis pipeline.}
\label{fig:instruction_synthesis}
\end{figure}

\subsection{Spectrogram Instruction Synthesis}
The captioning stage produces, for each spectrogram, a structured record that combines machine-readable signal metadata with a dense natural-language description of the scene. While these captions are technically accurate, they are (i) over-detailed, (ii) presented in a fixed order, and (iii) not formatted as instructions. Directly training on such monolithic captions makes optimization difficult and does not match the instruction-following setting we target for RF-GPT. To mitigate this, we (i) decompose caption information into several levels of difficulty and importance, and (ii) convert captions into diverse instruction–answer pairs using a text-only LLM. Specifically, we categorize caption information into five \textbf{information levels}:
\begin{itemize}
    \item \emph{Summary}: high-level description of the RF recording (e.g., ``a 5G NR downlink spectrogram containing several user data channels and synchronization signals'').
    \item \emph{Global visual}: global visual patterns that can be directly captured in the spectrogram e.g., presence of synchronization signal blocks, channel state information reference signal (CSI-RS)/ sounding reference signal (SRS) bursts, overall occupancy and structure.
    \item \emph{Global context}: spectrogram-level parameters that are not visually obvious but are known from metadata (e.g., channel bandwidth, subcarrier spacing, total time duration).
    \item \emph{Signal visual}: per-signal visual attributes (relative SNR and time/frequency occupancy).
    \item \emph{Signal context}: per-signal contextual parameters that are hard to infer visually, e.g., demodulation reference signal (DM-RS)/phase tracking reference signal (PTRS) settings for a given UE, detailed PDCCH/PUCCH structure.
\end{itemize}
Each instruction is constructed to present only a subset of these levels, so that the model focuses on a few related concepts per training example. In this work we mainly target summary and visual information (summary, global visual, signal visual) and leave fine-grained signal context, such as full DM-RS/PTRS or control resources set (CORESET) configurations, as future work. An example of a 5G NR spectrogram with its caption is illustrated in Fig. \ref{fig:fine_grained_caption_strategy}.

\textbf{Instruction synthesis pipeline.}
Starting from the caption JSON/JSONL files, we build an instruction-tuning dataset via the following steps (see Fig.~\ref{fig:instruction_synthesis} for an overview):

\begin{enumerate}
    \item \textbf{Load caption records:} We load the caption shards generated by captioning pipeline. Each record contains (i) structured signal metadata and (ii) one or more natural-language captions derived from that metadata.
    \item \textbf{Sample a task template:} For each record, the instruction builder samples a task template from a predefined library. Templates cover tasks such as signal counting, modulation or technology recognition and open-ended description.
    \item \textbf{Assemble the prompt:} Given a selected template, we construct a prompt for an LLM by combining a default \emph{system prompt} that describes the role of the model,  a task-specific \emph{user prompt}, and a selected subset of caption fields provided as hidden/internal context. These fields are drawn from the information levels above.
    \item \textbf{Generate with a text-only LLM:} The assembled prompt is sent to a powerful LLM, which produces a candidate response. The LLM is instructed to follow specific output formats depending on the template, such as plain labels (for classification), JSON snippets (for structured extraction), or short paragraphs (for explanations and open-ended QA).
    \item \textbf{Write instruction–answer pairs:} Successful pairs are recorded into dataset. Each entry contains the waveform identifier, the user-visible instruction, the target answer, the task type, and bookkeeping information. 
\end{enumerate}

This process converts heavy, deterministic captions into a large set of RF-grounded instruction–answer pairs that efficiently reflect how human experts query RF recordings. By controlling the task templates, information levels, and validation rules, we obtain a diverse and well-structured instruction dataset suitable for SFT. Beyond the scenarios around wireless technologies, to further enhance the diversity of modulation types, we also leverage TorchSig \cite{boegner2022torchsig} to generate diverse wide-band modulation classes. The workflow of captioning and instruction generation is similar. The resulting instruction–answer pairs constitute our RF instruction‑tuning dataset and also serve as the source for constructing the evaluation benchmarks in the following subsection.

\subsection{Benchmark Constructions}

Following the construction of the instruction datasets, we perform RF-grounded instruction tuning to transform a general-purpose VLM into an RFLM. To systematically evaluate the RFLM's RF understanding, we construct a comprehensive suite of benchmarks that tests the component recognition, quantity counting, and time/frequency relationship reasoning.

\subsubsection{Wide-Band Modulation Classification (WBMC)}

We build a modulation classification with 57 modulation classes leveraging TorchSig \cite{boegner2022torchsig} in a wideband setting with multiple, overlapping signals. In WBMC, each scene contains $2$–$5$ signals that may overlap in time and frequency. This benchmark evaluates the RFLM's ability to identify the modulations present in a wideband RF scene. We define three sub-tasks of increasing difficulty:

\begin{itemize}
    \item \textbf{Easy:} identify the set of modulation \emph{families} present in the spectrogram.
    \item \textbf{Medium:} identify the exact modulation class of each signal from a given candidate list.
    \item \textbf{Hard:} identify the exact modulation class of each signal from the full list of modulation classes.
\end{itemize}

Let a spectrogram contain $S$ signals with ground-truth modulation classes
$C = (c_1, \dots, c_S)$ \emph{according to their start time} and let the model output $M$ predicted classes
$C' = (\hat{c}_1, \dots, \hat{c}_{S'})$ \emph{using the same temporal ordering}. We define the per-example classification score with strict time order and number matching as,
\begin{equation}
    s^{\mathrm{CLS}} =
\begin{cases}
  \dfrac{1}{N} \displaystyle\sum_{i=1}^{N} \mathbbm{1}\{\hat{c}_i = c_i\}, & \text{if } S = S',\\[0.5em]
  0, & \text{otherwise}.
\end{cases} 
\end{equation}

In other words, if the predicted number of signals does not match the ground truth ($S \neq S'$), the answer receives a zero score. WBMC performance is reported as the average $s^{\mathrm{CLS}}$ over the test set for each difficulty level. For each level, we generate 2000 visual question–answer (VQA) pairs from RF samples with the co-channel overlap probability set to zero, so that overlapping signals do not affect the modulation classification results severely\footnote{As noted in the TorchSig documentation, setting the co-channel overlap probability to zero does not completely eliminate overlapping signals.}. Time/frequency overlap analysis is instead evaluated separately by the benchmark introduced below.

\subsubsection{Wide-Band Overlap Detection (WBOD)}

WBOD evaluates whether an RFLM can reason about how RF signals overlap in time and frequency. From the metadata we take each signal's time support $T_i=[t_i^{\mathrm{start}},t_i^{\mathrm{end}}]$ and frequency support $F_i=[f_i^{\mathrm{low}},f_i^{\mathrm{high}}]$, and label a pair by whether their supports intersect in time, in frequency, in both, or in neither, giving four overlap types: \emph{neither}, \emph{time-only}, \emph{frequency-only}, and \emph{both}.

\textbf{Easy (global overlap):} a single scene-level label, set to the strongest overlap present. It is \emph{both} if any pair overlaps on both axes or if a time-only pair and a frequency-only pair co-occur, \emph{time-only} or \emph{frequency-only} if only one axis is ever involved, and \emph{neither} otherwise.

\textbf{Medium (pairwise overlap):} the instruction names two signals and the model outputs one of the four types; accuracy is the fraction of pairs predicted correctly.

\textbf{Hard (overlap strength):} the model also rates how strongly a pair $(A,B)$ overlaps on each axis, using the time- and frequency-overlap ratios
\begin{equation}
r_t(A,B) = \frac{|T_A \cap T_B|}{|T_A \cup T_B|}, \qquad
r_f(A,B) = \frac{|F_A \cap F_B|}{|F_A \cup F_B|}.
\end{equation}
Each ratio is quantized into four levels, \emph{none} below $0.01$, \emph{slightly} up to $0.3$, \emph{considerably} up to $0.6$, and \emph{almost fully} beyond. The label pairs the time and frequency levels and counts as correct only when both match.

WBOD-Medium and Hard consider only \textbf{adjacent} signals in the time- or frequency-sorted order, which targets the locally relevant interactions and avoids a combinatorial find-any-pair search that would entangle WBOD with WBMC. As in WBMC, each difficulty level has 2000 VQAs at co-channel overlap probability $0.6$ and SNR from $10$ to $50$\,dB.

\subsubsection{Wireless Technology Recognition (WTR)}

For WTR, an RFLM is required to identify the wireless technology present in a spectrogram and, when applicable, the link direction (downlink vs.\ uplink). The label space consists of technology–direction pairs derived directly from the waveform generator configuration for technologies including NR, LTE, and WLAN. Performance is reported as standard top-1 classification accuracy over the test set. The test set contains 1000 samples for each of DVB-S2, Bluetooth, UMTS, and LTE, 1000 NR downlink and 1000 NR uplink signals, and 1000 WLAN-AX (11ax) and  WLAN-BE (11be) signals.

\subsubsection{WLAN Number of Users Counting (WNUC)}

WNUC evaluates an RFLM’s ability to estimate the number of simultaneous users present in a wideband WLAN spectrogram. Each sample is labeled with the ground-truth user count $U$, denotes the number of distinct WLAN users that are active within the time–frequency window represented by the spectrogram. This can be obtained from the WLAN configuration metadata. Instead of predicting the exact number of users, we formulate an easier classification task by grouping user counts into fixed-size intervals. Hence, we define three difficulty levels:

\begin{itemize}
    \item \textbf{Easy (15-user buckets):} classification over coarse bins of size $B=15$. The label is the bucket interval $[s,e]$ where
    \begin{equation}
    s = \left\lfloor \frac{U-1}{B} \right\rfloor B + 1,\quad e = s + B - 1.
    \end{equation}
    For example, with $B=15$ and $U=17$, we obtain $[s,e] = [16,30]$.
    
    \item \textbf{Medium (10-user buckets):} identical to Easy but with $B=10$. For instance, with $B=10$ and $U=17$, we obtain $[11,20]$.
    
    \item \textbf{Hard (numeric with rounding):} the model outputs a single integer $\hat{U}$. If $U<10$, the correct answer is the exact count. If $U\ge 10$, the correct answer is the nearest multiple of 10 (standard rounding):
    \begin{equation}
      \hat{U}^* =
      \begin{cases}
      U, & U<10,\\[0.3em]
      10\cdot \mathrm{round}(U/10), & U\ge 10.
      \end{cases}
    \end{equation}
\end{itemize}

For easy and medium levels, we report bucket-level accuracy (predicted interval matches the true interval). For hard level, we report the fraction of examples where the predicted integer equals $\hat{U}^*$. Our benchmark test set consists of 1000 WLAN-AX samples and 1000 WLAN-BE samples.

\subsubsection{New Radio Information Extraction (NRIE)}

NRIE focuses on interpreting 5G NR spectrograms because (i) NR is a dominant cellular standard and (ii) its time--frequency structure encodes rich PHY-layer information. Each NRIE sample presents a single NR spectrogram (e.g., a 10\,ms window) and asks the model to answer \emph{one} question about a specific attribute. The evaluated attributes include the number of UEs, subcarrier spacing (SCS) in kHz, SSB pattern (DL only), CSI-RS count (DL only), and SRS count (UL only).

NRIE therefore evaluates a model’s ability to extract key NR parameters from visual evidence alone, with some questions conditioned on link direction (DL/UL). For all NRIE tasks, we report exact-match accuracy between the parsed model output and the ground-truth label. Our test set for NRIE consists of 1000 DL samples and 1000 UL samples. More fine-grained NR features, such as PTRS and per-UE DMRS patterns or detailed PDCCH/PUCCH and CORESET configurations, will be considered in future work. It is worth noting that the hard-level benchmarks are designed to require a certain degree of reasoning capability, such that they can serve as meaningful indicators for future RFLMs.

\section{Evaluations}
\label{sec:method}

\subsection{Experimental Settings}

We build RF-GPT on top of Qwen2.5-VL with 3B and 7B parameters, denoted as RF-GPT-3B and RF-GPT-7B, respectively, each of which is fine-tuned on the synthetic RF instruction dataset. As baselines, we evaluate the off-the-shelf {Qwen2.5-VL-3B-Instruct}, {Qwen2.5-VL-7B-Instruct}, and GPT-5 without any RF-specific fine-tuning. We fine-tune both models, RF-GPT-3B and RF-GPT-7B, with 3 epochs using AdamW \cite{kingma2015adam} with learning rate $2\times 10^{-4}$, global batch size 256, 5\% warm-up and a cosine decay schedule. All models are trained with mixed precision (BF16) in PyTorch on 8 NVIDIA H200 GPUs with 140 GB video memory. For all experiments, complex baseband IQ sequences are converted into spectrograms using an STFT with a Blackman window (no centering, FFT shift along the frequency axis), FFT size 512, window length 512 samples, and hop size 512 samples. The magnitude spectrogram is converted to dB, clipped to a fixed dynamic range, and resized to $512\times 512$ pixels, the vision encoder processes non-overlapping $14\times 14$ patches, yielding 1296 visual patches, merged by the Qwen2.5-VL processor into 324 image tokens per spectrogram. All RF instructions are generated following the procedure explained in Sec. \ref{sec:rf_grounding} with GPT-OSS-120B~\cite{agarwal2025gptoss}. In total, our synthetic pipeline generates approximately 120,000 distinct RF scenes and 0.625 million RF‑grounded instruction–answer pairs. Train, validation, and test scenes occupy disjoint directories with unique waveform identifiers under deterministic per-technology seeding, so the benchmark test split is statistically independent of training.

\subsection{General-Purpose VLMs Show Very Limited RF-Specific Understanding}

\begin{figure}[t!]
\centering
\begin{mdframed}[linecolor=black, linewidth=2pt, roundcorner=10pt]
\textbf{An Example of 5G NR Spectrogram with One UE}
\hrule height 0.5pt

\vspace{0.08in}
\textbf{User:} Describe what do you observe in this RF scene.

\vspace{0.02in}
\begin{center}
  \includegraphics[width=0.4\linewidth]{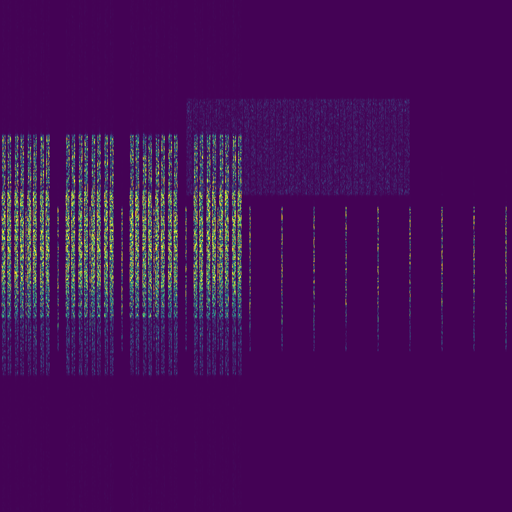}
\end{center}

\vspace{0.02in}
\textbf{Qwen-2.5-VL-7B-Instruct}: This spectrogram displays a variety of radio-frequency (RF) signals. In the central region, there are several prominent vertical lines, suggesting strong, narrowband transmissions that could be from specific transmitters such as amateur radio stations, radar systems, or other point-to-point communication devices. To the right, there are fainter, more dispersed lines, which might indicate lower power transmissions or background noise...

\vspace{0.05in}
\textbf{RF-GPT-7B}: The spectrogram shows a single‑UE \textbf{5G‑NR} downlink capture in the \textbf{FR2} band. A SSB is present, following the \textbf{Case D} transmission pattern, and the PDCCH is also active. \textbf{No CSI‑RS} traces are visible, confirming that CSI‑RS transmission is disabled for this capture. The data payload occupies one continuous block of PRBs. The occupied bandwidth is moderate. In the time dimension the burst is \textbf{short}, lasting only a few symbols before the signal fades...  
\end{mdframed}
\caption{Comparison between Qwen2.5-VL-7B-Instruct and RF-GPT on a 5G DL spectrogram with only one UE.}
\label{fig:mcq_vqa_rationale_demo}
\end{figure}

To further motivate RF-GPT framework and showcase its potential, we first study whether existing general-purpose VLMs possess any useful RF prior. First, we compare the response of RF-GPT-7B and its base model Qwen2.5-VL-7B-Instruct for a concrete 5G DL signal in Fig. \ref{fig:mcq_vqa_rationale_demo}. We can observe that the response of general-purpose VLM is shallow and does not provide any meaningful information. On the other hand, RF-GPT provides grounded and accurate answers including SSB patterns, non-existence of CSI-RS, number of UEs, and their time/frequency occupancy.

\textcolor{black}{To put the per-benchmark numbers in the following subsections on a common reference frame, we additionally report two naive baselines per benchmark (Table~\ref{tab:naive_baselines}): a random-guess baseline that samples uniformly over the candidate label set, with WBMC sampling over the candidate vocabulary; and a majority-class baseline that always predicts the most frequent label observed in the test split, jointly over all output components for WTR and WBOD-Hard. The general-purpose VLM scores reported below sit close to or below these naive references on most tasks, indicating \emph{very limited RF-specific understanding under our benchmark setting} and motivating RF-grounded instruction tuning, as done in RF-GPT.}

\begin{table}[t!]
\centering
{
\captionsetup{font={color=black},labelfont={color=black}}
\caption{Random-guess and majority-class baseline accuracies (\%), averaged across difficulty splits.}
\label{tab:naive_baselines}
\small
\setlength{\tabcolsep}{6pt}
\renewcommand{\arraystretch}{1.15}
\begin{tabular}{lcc}
\hline
Benchmark & Random & Majority \\
\hline
WTR       &  5.27 & 12.70 \\
NRIE      & 23.66 & 29.00 \\
WBOD      & 19.44 & 41.33 \\
WBMC      &  1.11 &  3.30 \\
WNUC      &  9.60 & 27.32 \\
\hline
\end{tabular}
}
\end{table}

\subsection{Benchmark Results: RF-GPT vs. General-purpose Models}

Here, we present the performance of RF-GPT on the proposed benchmarks and compare it against general-purpose VLMs. Across all tasks, RF-GPT substantially outperforms the baseline models, validating the effectiveness of spectrogram-based RF grounding.

\begin{figure}[t!]
\centering
\includegraphics[width=0.9\linewidth]{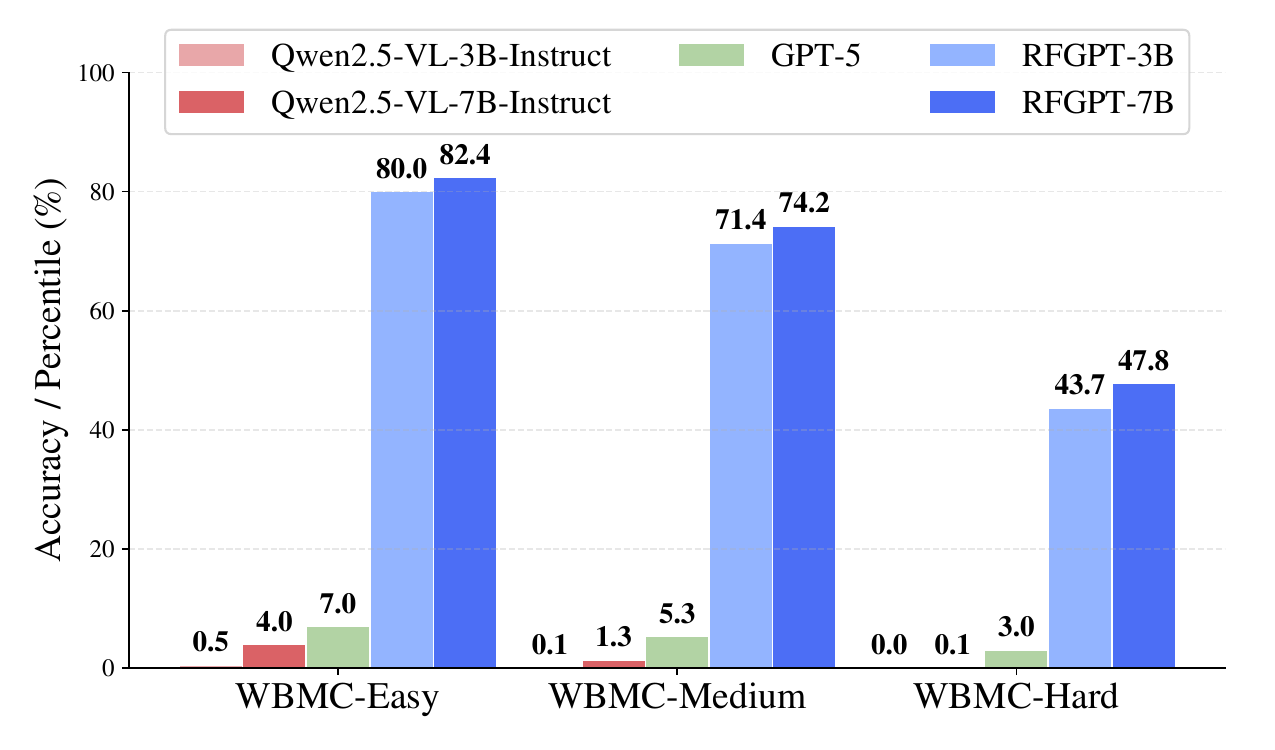}
\vspace{-3mm}
\caption{Benchmark scores on the wide‑band modulation classification (WBMC) tasks.}
\label{fig:wbmc_results}
\end{figure}

\textbf{WBMC:}  Fig.~\ref{fig:wbmc_results} presents the WBMC results. General-purpose VLMs achieve at most $7\%$ accuracy on the easy task and close to $0\%$ on the hard task, with an average score below $2\%$. These extreme low scores are due to our strict time-order awareness and number matching criteria. In fact, most of general-purpose VLMs fail to predict the number of signals presented in the spectrogram. In contrast, RF-GPT-3B reaches $80.0\%$ / $71.4\%$ / $43.7\%$ accuracy on Easy / Medium / Hard, and RF-GPT-7B further improves the accuracy performance to $82.4\%$ / $74.2\%$ / $47.8\%$, with respect to the same levels. The average score increases largely after RF-grounded finetuning of both 3B and 7B models. Importantly, RF-GPT correctly identifies the number of signals in around $ 98\%$ of the cases, while general VLMs almost fail all the cases. These results indicate that RF-GPT can both estimate the number of active signals and correctly identify their modulation types in complex wideband input.

\begin{figure}[t!]
\centering
\includegraphics[width=0.9\linewidth]{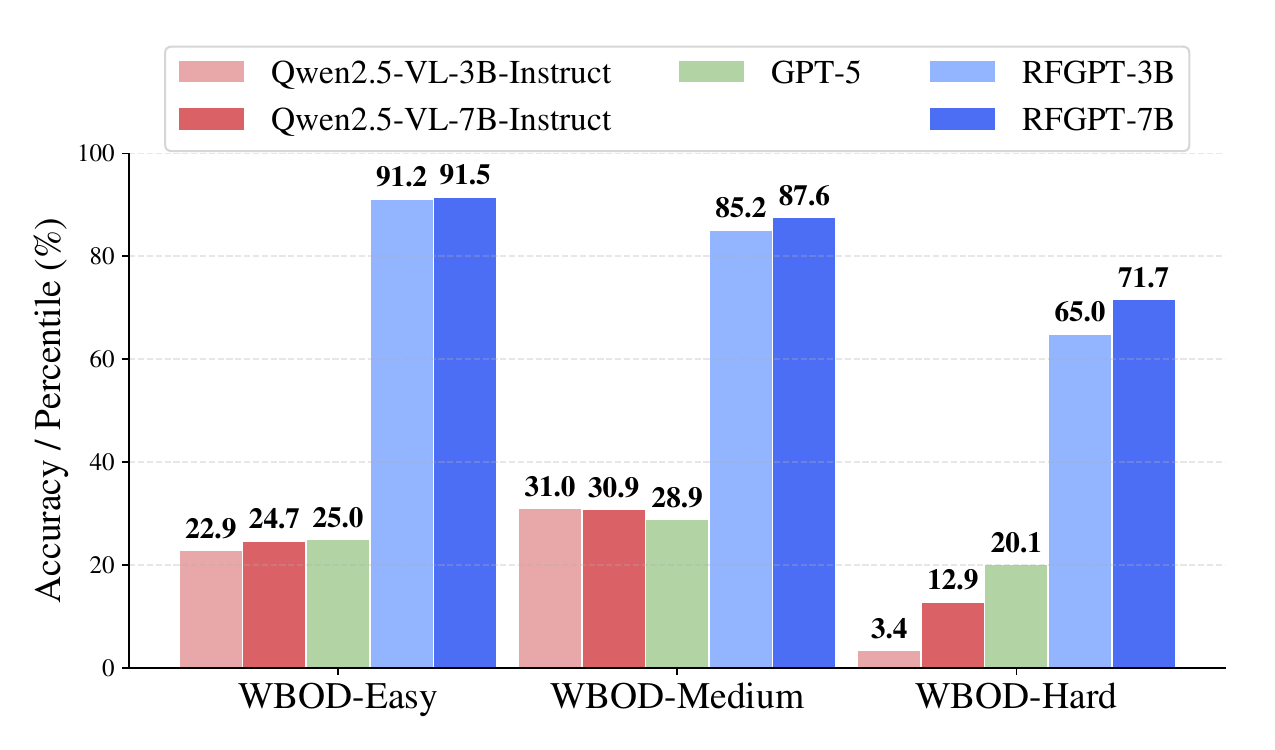}
\vspace{-4mm}
\caption{Benchmark scores on the wide‑band overlap detection (WBOD) tasks.}
\label{fig:wbod_results}
\end{figure}

\textbf{WBOD:} WBOD results are shown in Fig.~\ref{fig:wbod_results}. General-purpose VLMs perform marginally above chance. Their easy-level accuracies range from approximately $23$–$25\%$, medium-level accuracies remain around $30\%$, and hard-level time–frequency overlap classification drops to $5$–$12\%$, resulting in overall average scores of roughly $11$–$16\%$. These numbers are partly influenced by a biased prediction pattern in which the models frequently select the ``none'' label, even though only a small fraction of signal pairs exhibit no overlap in either time or frequency. In contrast, RF-GPT-3B achieves $91.2\%$ (Easy), $85.2\%$ (Medium), and about $65.0\%$ accuracy on the hard joint time–frequency overlap task. RF-GPT-7B further improves the hard-task accuracy to $71.7\%$, with similar or slightly better performance on the easy and medium tasks. These results indicate that RF-GPT can reason coherently about whether signals overlap in time and frequency and to what extent, whereas generic VLMs produce essentially random overlap labels.

\begin{figure}[t!]
\centering
\includegraphics[width=0.9\linewidth]{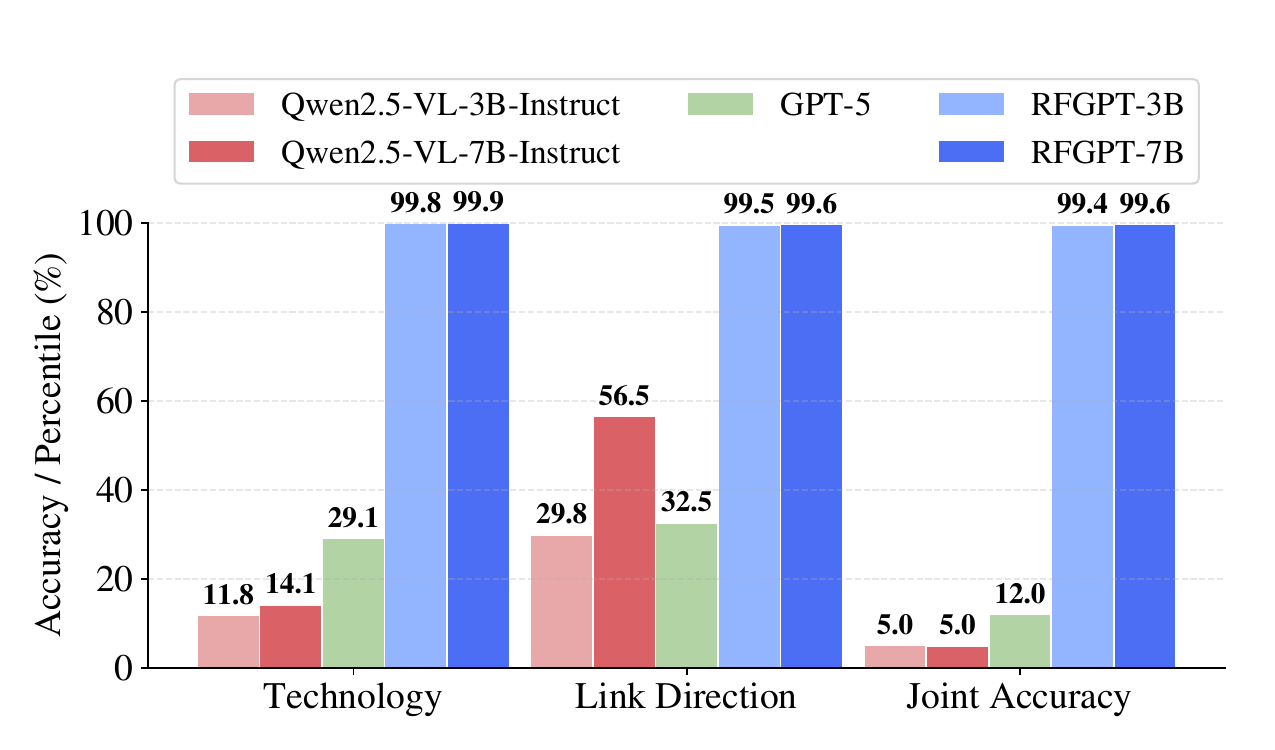}
\vspace{-4mm}
\caption{Benchmark results for wireless technology recognition (WTR) benchmark.}
\label{fig:wtr_results}
\end{figure}

\textbf{WTR:} Fig.~\ref{fig:wtr_results} summarizes WTR performance. Qwen2.5-VL series achieve a joint technology+link accuracy of only $5.01\%$ (3B) and $4.98\%$ (7B), with highly inconsistent behavior across technologies. For example, {Qwen2.5-VL-3B-Instruct} predicts LTE reasonably often but almost never identifies NR, UMTS, WLAN, or DVB-S2 correctly, and its link-direction predictions are heavily biased (e.g., always ``DL''). GPT-5 does not produce any consistent correct predictions under our parsers. In contrast, RF-GPT-3B achieves $99.42\%$ joint accuracy, with almost perfect per-technology and per-link scores ($\geq 99.5\%$ on most classes). RF-GPT-7B further improves to $99.64\%$ joint accuracy and close to $100\%$ on most individual technologies and link directions. This shows that RF-GPT has learned a highly reliable RF taxonomy across heterogeneous standards.

\begin{figure}[t!]
\centering
\includegraphics[width=1.0\linewidth]{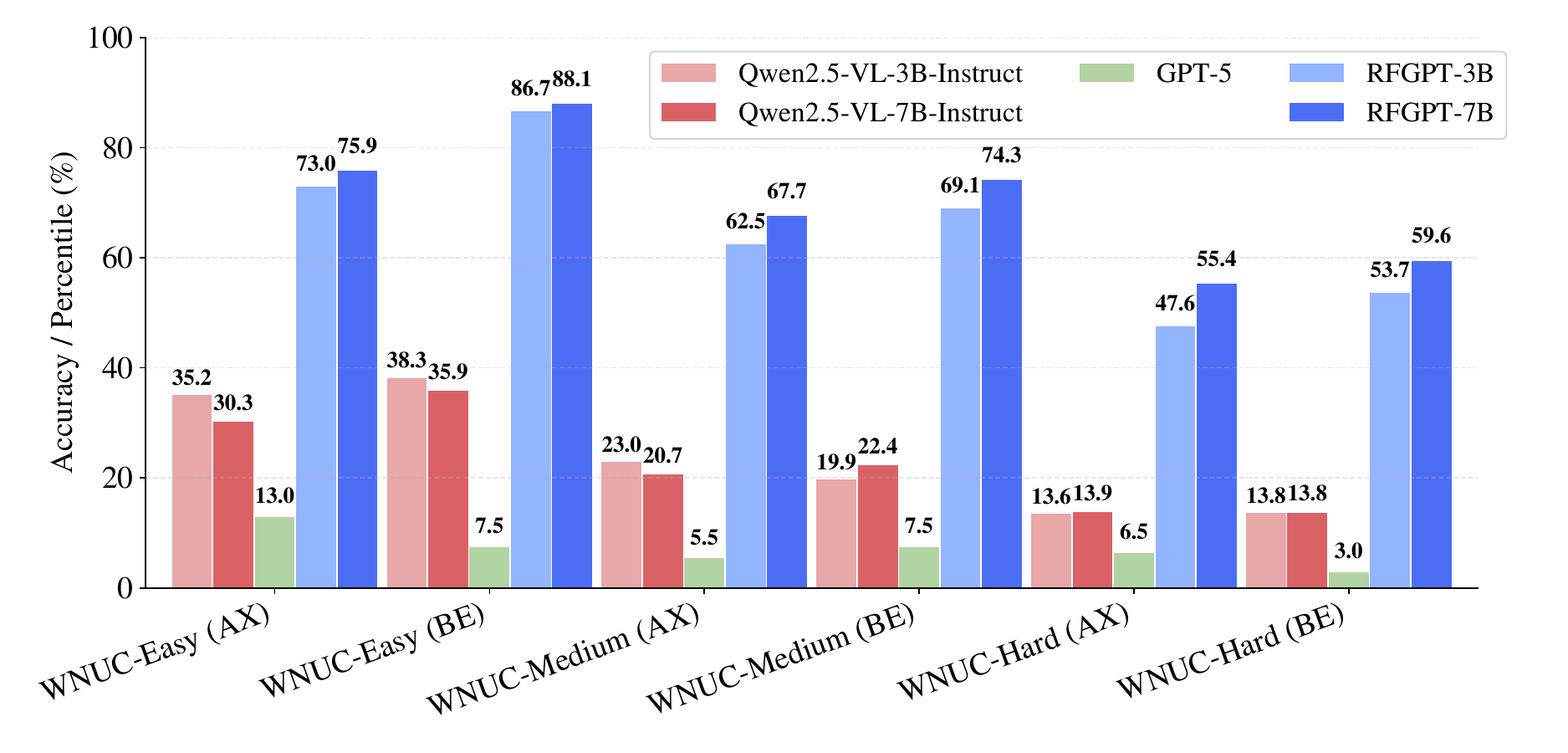}
\vspace{-8mm}
\caption{Benchmark scores on the WLAN number of users counting (WNUC) tasks for 802.11ax and 802.11be.}
\label{fig:wnuc_results}
\end{figure}

\textbf{WNUC:} WNUC results are presented in Fig.~\ref{fig:wnuc_results}. General-purpose VLMs achieve average accuracies of $23.97\%$ (3B) and $22.83\%$ (7B) across all difficulty levels and both WLAN standards (11ax/11be). They struggle particularly on the hard (numeric) task, where accuracies are around $13$–$14\%$. After RF grounding, the same backbones achieve much higher scores, where the 3B model attains an average of $65.43\%$, and the 7B model reaches $70.17\%$. In particular, both model variants exhibit consistent higher performance in the 802.11be standard compared to 11ax. For RF-GPT-3B, we observe absolute accuracy gains of $13.7\%$, $6.6\%$, and $6.1\%$ for the easy, medium, and hard difficulty levels, respectively. Similarly, for the larger RF-GPT-7B model, the gains are $12.2\%$, $6.6\%$, and $4.2\%$, respectively. {This performance difference stems from the different resource mapping strategies used during data generation. While 11be samples typically assign a single user per Resource Unit (RU), 11ax samples frequently involve multiple users sharing a single RU through MU-MIMO.} The latter creates significant visual overlap in the time-frequency plane, masking the individual user signatures and making the counting task more challenging for a vision-based backbone compared to the spatially distinct OFDMA patterns in 11be. This demonstrates that RF-GPT can infer approximate user counts directly from OFDMA/MU-MIMO structures in WLAN spectrograms.

\begin{figure}[t!]
\centering
\includegraphics[width=1.0\linewidth]{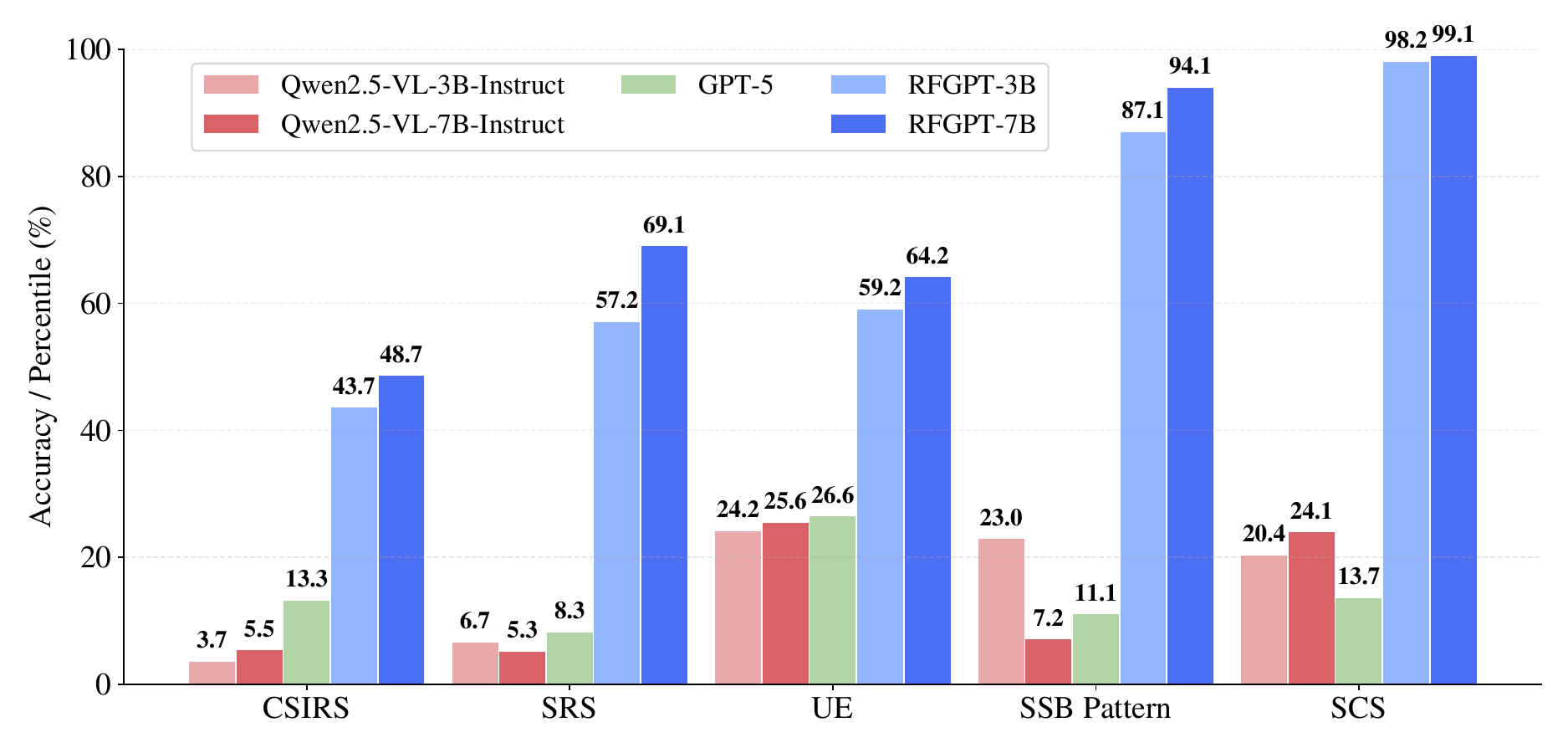}
\vspace{-6mm}
\caption{Benchmark results for the New Radio information extraction (NRIE) benchmark.}
\label{fig:nrie_results}
\end{figure}

\textbf{NRIE:} NRIE results are shown in Fig.~\ref{fig:nrie_results}. General-purpose VLMs perform poorly on NR-specific attributes, with average accuracies around $20\%$ and substantial variation across tasks (e.g., slightly better on UE count, but weak on SSB pattern and SCS). After RF grounding, RF-GPT-3B reaches an average accuracy of $72\%$, with near-perfect SCS and SSB pattern recognition and moderate performance on UE number estimation, CSI-RS, and SRS resources. RF-GPT-7B achieves similar or improved performance on most tasks (notably higher SRS and UE estimation), confirming that larger models can extract richer NR structure from spectrograms. Overall, NRIE shows that RF-GPT is not limited to generic modulation or technology recognition, but it can also recover protocol-specific parameters that are crucial for 5G NR analysis.

Taken together, these results show that RF-GPT consistently turns general-purpose VLMs, which exhibit very limited RF-specific understanding under our benchmark setting, into strong RF foundation models that perform well across a diverse suite of benchmarks, including modulation classification, overlap analysis, technology recognition, number of users estimation, and NR-specific attribute extraction.

\begin{figure}[t!]
\centering
\includegraphics[width=0.75\linewidth]{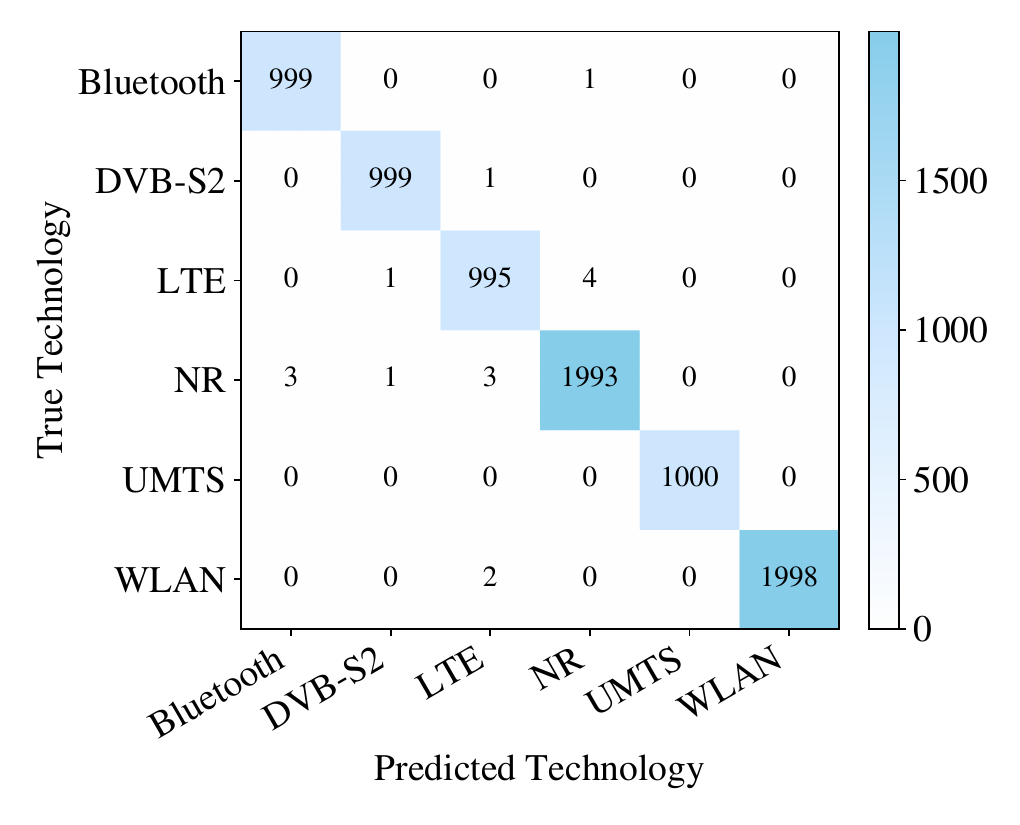}
\caption{WTR confusion matrix on RF-GPT-7B.}
\label{fig:wtr_confusion_matrix}
\end{figure}

\begingroup
\subsection{Failure-Case Analysis}
\label{sec:failure_analysis}

\textcolor{black}{\textbf{WBMC.} Table~\ref{tab:wbmc_failure_breakdown} classifies each test sample as fully correct, count error (under/over-predict), wrong temporal ordering, or per-signal class confusion. Counting and ordering errors are both small, with a combined fraction $\le 5\,\%$ at every difficulty; the dominant error mode is per-signal class confusion, which grows from $43.10\,\%$ at Easy to $87.85\,\%$ at Hard. The top confusions are PSK predicted as QAM at Easy (visually similar families) and ofdm-64 vs ofdm-72 at Hard, fine-grained variants that differ only in subcarrier count. This pattern is consistent with the resolution ablation in Fig.~\ref{fig:WBMC_image_resolution_ablation} and suggests that further WBMC-Hard gains require either higher-resolution spectrograms or phase-aware representations.}

\begin{table}[!htbp]
\centering
\captionsetup{font={color=black},labelfont={color=black}}
\caption{WBMC error decomposition on RF-GPT-7B.}
\label{tab:wbmc_failure_breakdown}
\small
\setlength{\tabcolsep}{4pt}
\renewcommand{\arraystretch}{1}
\begin{tabular}{lccc}
\hline
Failure mode                 & Easy   & Medium & Hard   \\
\hline
Fully correct                & 53.80  & 40.60  &  9.20  \\
Count error (under-predict)  &  2.00  &  2.05  &  2.15  \\
Count error (over-predict)   &  0.30  &  0.20  &  0.50  \\
Wrong temporal ordering      &  0.80  &  4.90  &  0.30  \\
Per-signal class confusion   & 43.10  & 52.25  & 87.85  \\
\hline
\end{tabular}
\end{table}

For WTR, the joint technology and link-direction confusion matrix in Fig.~\ref{fig:wtr_confusion_matrix} is essentially diagonal: of the $8000$ joint test cases only $18$ are misclassified, for $99.78\,\%$ joint accuracy. The off-diagonal entries are concentrated in two cell pairs, NR DL vs NR UL and WLAN DL vs WLAN UL, so the residual confusion reflects fine-grained DL/UL slot-pattern ambiguity within a technology rather than technology-level mis-recognition.
\endgroup

\begingroup
\subsection{Deployment: Latency and Memory Analysis}
\label{subsec:deployment}

We measure inference latency and peak GPU memory of RF-GPT-3B/-7B on a single NVIDIA H200, mapping each $512\!\times\!512$ spectrogram to 324 image tokens and greedily decoding 64 output tokens, with medians over 30 cold queries. End-to-end raw-IQ-to-answer cost on the 3B model is $\sim 293$\,ms: time-to-first-token (TTFT) is $\sim 48$\,ms, dominated by the vision encoder at $\sim 32$\,ms, and decode runs at $3.07$\,ms per token. The 7B model adds $\sim 112$\,ms, almost entirely from slower decode at $4.79$\,ms per token, while TTFT rises only marginally to $\sim 50$\,ms since the vision encoder is shared. The dominant CPU overhead is the $\sim 51$\,ms IQ-to-image preprocessing, a natural target for further optimization. The 7B fits on a single $16$\,GB consumer GPU. At response times in the hundreds of milliseconds, RF-GPT is suited to interactive RF analysis, spectrum monitoring, and AI-native operator dialogue rather than sub-millisecond PHY control loops.

\begin{table}[t!]
\centering
\caption{Inference latency and peak HBM on NVIDIA H200.}
\label{tab:deployment_latency}
\small
\setlength{\tabcolsep}{6pt}
\renewcommand{\arraystretch}{1}
\begin{tabular}{lcc}
\hline
Component                       & RF-GPT-3B  & RF-GPT-7B  \\
\hline
STFT (CPU)                      & 0.65\,ms   & 0.95\,ms   \\
Image preproc.\ (CPU)           & 51.34\,ms  & 51.78\,ms  \\
Vision encoder (GPU)            & 31.64\,ms  & 32.03\,ms  \\
LLM prefill (GPU, derived)      & 16.00\,ms  & 18.05\,ms  \\
LLM decode (GPU, per token)     & 3.07\,ms   & 4.79\,ms   \\
\hline
End-to-end (1 img + 64 tok)     & 292.93\,ms & 404.71\,ms \\
\hline
Weights resident (HBM)          & 7.03\,GB  & 15.48\,GB \\
Peak HBM (wgt + act + KV)       & 7.09\,GB  & 15.59\,GB \\
\hline
\end{tabular}
\end{table}
\endgroup

\subsection{Ablation Studies}

\begin{figure}[t!]
\centering
\includegraphics[width=0.8\linewidth]{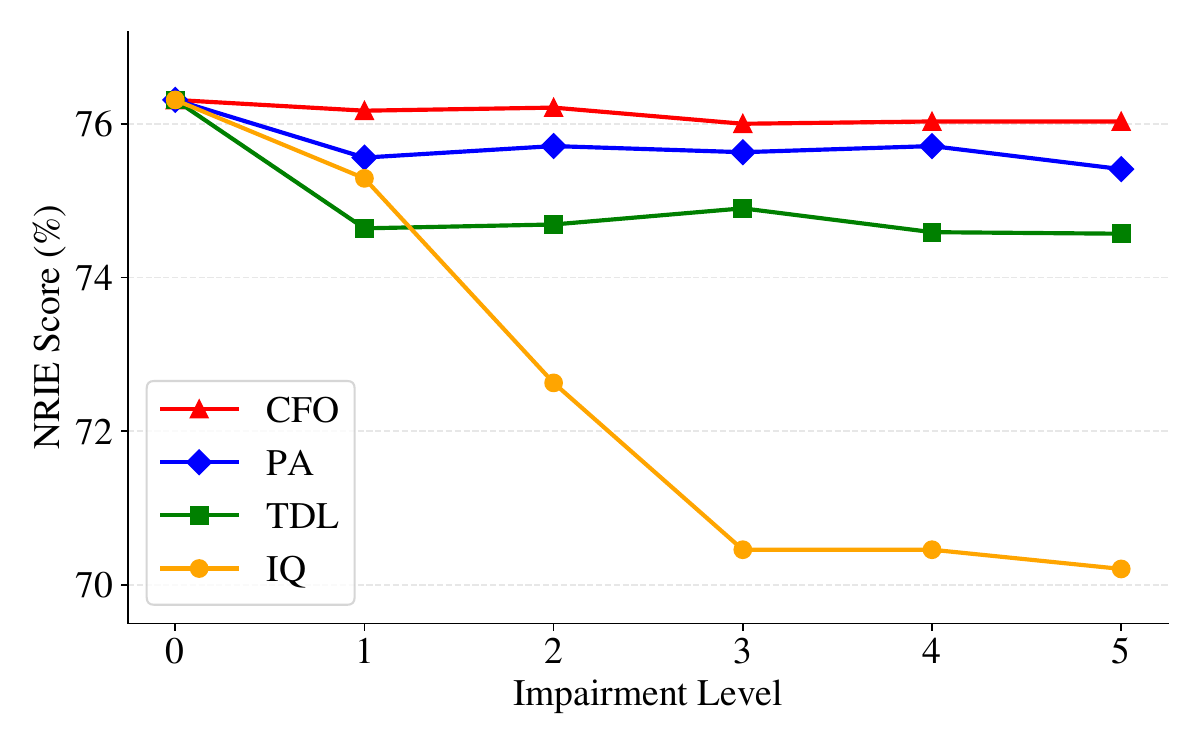}
\vspace{-4mm}
\caption{NRIE ablation vs. impairment level of IQ imbalance (IQ), power amplifier nonlinearity (PA), tapped delay line (TDL), and carrier frequency offset (CFO).}
\label{fig:NRIE_impairment_sensitivity}
\end{figure}

\textbf{Robustness against impairments.}
We start by evaluating the robustness of RF-GPT to common RF impairments. For impairment-sweep experiments, we vary one impairment family at a time, including in-phase/quadrature (IQ) imbalance, power amplifier (PA) nonlinearity, carrier frequency offset (CFO), or tapped delay line (TDL) channel models, while reusing a common clean waveform as a base. We introduce a normalized impairment level $\lambda \in \{0.2, 0.4, 0.6, 0.8, 1.0\}$, with $\lambda=0$ corresponding to the clean baseline. During training, we include \textbf{only data samples with relatively mild impairments} ($\lambda \leq 0.3$), so that robustness to larger distortions must arise from generalization rather than direct exposure during training. Fig.~\ref{fig:NRIE_impairment_sensitivity} reports NRIE performance of RF-GPT-7B under increasing impairment severity for CFO, TDL, PA, and IQ imbalance. Starting from a clean score of \(76.31\%\), CFO and PA sweeps cause only minor degradation and the curves remain nearly flat, while TDL introduces a moderate drop. In contrast, IQ imbalance produces the largest and nearly monotonic performance decline, dropping by about \(6.1\) percentage points from \(75.29\%\) at the lowest nonzero level to \(70.21\%\) at the highest level. The ranking matches how each impairment alters spectrogram structure: CFO, PA, and TDL largely preserve the coarse NR/OFDM layout, whereas IQ imbalance introduces image-frequency cross-leakage and mirror-like duplicated patterns that directly corrupt the counting and reference-signal cues NRIE relies on, making it the most destructive. This also explains the within-NRIE spread in the main results (Fig.~\ref{fig:nrie_results}): the categorical attributes SCS and SSB pattern are essentially saturated, whereas the counting questions on UE, SRS, and especially the sparse, data-layer-overlapped CSI-RS are markedly harder. The mirror effect of IQ imbalance duplicates and shifts these sparse reference-signal regions, severely degrading counting tasks while leaving categorical attributes largely intact. \textcolor{black}{These results indicate that RF-GPT generalizes to impairment levels beyond the mild conditions encountered during training, although its performance gradually declines as the impairments become more severe. This evaluation considers different impairment levels under a fixed waveform configuration and therefore does not assess generalization across waveform settings. Evaluating RF-GPT on unseen bandwidths, numerologies, and other operating points remains an important direction for future work.}

\textbf{Comparison with CNN/Transformer baselines.}
To compare RF-GPT with strong non-LLM baselines, we implement spectrogram-based CNN and Transformer models such as EfficientNet~\cite{tan2019efficientnet} and ViT~\cite{dosovitskiy2020image} for the NRIE benchmark, each with a shared backbone and separate task-specific heads per NRIE attribute. These task-specific customizations underscore the narrow generalization of conventional deep neural networks compared to a unified instruction-following RFLM.

\begin{figure}[t!]
\centering
\includegraphics[width=0.8\linewidth]{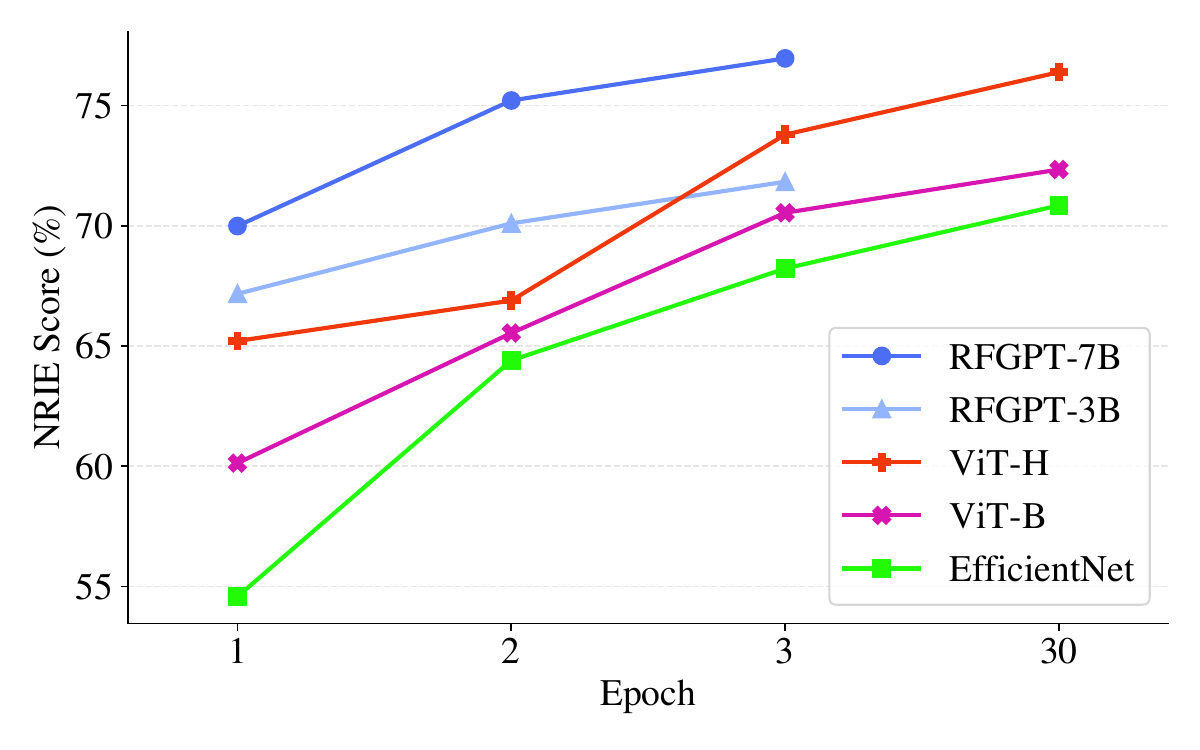}
\vspace{-4mm}
\caption{NRIE ablation comparing RF-GPT and Transformer/CNN baselines across epochs.}
\label{fig:NRIE_cnn_ablation}
\end{figure}

Fig.~\ref{fig:NRIE_cnn_ablation} compares NRIE performance across training for RF-GPT and the CNN/Transformer baselines. RF-GPT-7B leads throughout, reaching \(76.96\%\) after only 3 epochs and slightly exceeding the strongest baseline, ViT-H, which needs 30 epochs to reach \(76.39\%\); RF-GPT-3B is already competitive with ViT-B by epoch 3. This data efficiency reflects the NRIE task structure: jointly inferring heterogeneous attributes benefits more from a unified language-conditioned model than from fixed closed-set multi-head classifiers. In contrast to multi-head CNN or Transformer classifiers that require separate architectures and supervision for each NRIE task, RF-GPT relies on a single token-level objective and handles different NR attributes through language instructions, without modifying the architecture.

\textcolor{black}{\textbf{Comparison with RF-specific multi-signal AMC baselines.} Single-signal CNN/Transformer AMC architectures do not apply to WBMC because each scene contains $2$--$5$ co-existing signals that must be localized before labelling; published multi-signal AMC instead reformulates the problem as \emph{joint signal detection and classification}, as in Vagollari \emph{et al.}~\cite{vagollari2021jointDetectClassif}, HIFI-YOLO~\cite{hifi_yolo_2025}, and RF-YOLO~\cite{rf_yolo_2025}. We therefore train two RF-specific baselines on the same 40\,000-scene WBMC train split as RF-GPT (Table~\ref{tab:wbmc_baselines}). The first is an Oracle-localization $+$ ResNet18 AMC head, which assumes perfect per-signal detection and upper-bounds the classification step. The second is a stock YOLOv8s detect-and-classify pipeline~\cite{ultralytics_yolov8}, a literature-faithful member of the HIFI-YOLO/RF-YOLO family, scored on the same rule of WBMC after sorting detections. YOLOv8s outperforms RF-GPT-7B at the family level on Easy, $89.49\%$ against $82.41\%$, whereas the picture inverts at the fine-grained level: RF-GPT-7B leads substantially on Medium and Hard, $74.24\%$ and $47.77\%$ against $24.77\%$ and $24.77\%$. Both prompt-blind baselines score identically at Medium and Hard because the Medium/Hard distinction is encoded by whether the prompt supplies a per-scene candidate list (Sec.~III), which a vision-only classifier cannot use. The fine-grained inversion itself is driven by YOLO's count-accuracy collapsing from $92.75\,\%$ at family-level to $35.10\,\%$ at fine-grained when $57$ visually-similar OFDM/PSK/QAM variants must be separated. The oracle ResNet18 ($63.61\,\%$ fine-grained) caps the classification step when detection is solved; RF-GPT jointly handles counting, ordering, and classification through a single instruction-following interface, sitting between the classification-only oracle and the detect-and-classify YOLO on Medium/Hard.}

\begin{table}[t!]
\centering
\captionsetup{font={color=black},labelfont={color=black}}
\caption{WBMC baselines vs RF-GPT.}
\label{tab:wbmc_baselines}
\small
\setlength{\tabcolsep}{8pt}
\renewcommand{\arraystretch}{1}
\begin{tabular}{lccc}
\hline
Method                  & Easy  & Medium & Hard  \\
\hline
Oracle-Loc + ResNet18   & 99.43 & 63.61  & 63.61 \\
YOLOv8s det+cls         & 89.49 & 24.77  & 24.77 \\
RF-GPT-3B               & 80.00 & 71.40  & 43.69 \\
RF-GPT-7B               & 82.41 & 74.24  & 47.77 \\
\hline
\end{tabular}
\end{table}

\begingroup
\textbf{Design-choice ablations.}
We ablate four design choices on RF-GPT-7B in Table~\ref{tab:ablations_design}: the trained submodule, pseudo-RGB vs grayscale spectrograms, instruction-set size of $25$, $50$, or $100\,\%$, and varied task-templated prompts vs a single ``Describe this RF spectrogram.'' prompt. The submodule variants are LLM-only, vision-encoder-only, and full-finetuning. Full-finetuning leads on every benchmark family except NRIE, where it falls below LLM-only; vision-encoder-only is above LLM-only on perception-heavy benchmarks but at or below it on integer-output and tech-recognition tasks. We keep LLM-only as the default because it preserves the pretrained multimodal foundation, whereas tuning the vision encoder to spectrogram-only statistics would erode the general image understanding needed for future cross-modality extension. Dropping the viridis colormap degrades every task, so the pretrained encoder reads the colormap as additional RF information. Accuracy rises monotonically with instruction-set size, with diminishing returns past $50\,\%$. The naive uniform-prompt baseline collapses on nearly every task, with WTR falling from $99.64\%$ to $38.74\%$, confirming the contribution of our task-templated instruction synthesis. 
\endgroup

\begin{table}[t!]
\centering
\caption{Design-choice ablations on RF-GPT-7B.}
\label{tab:ablations_design}
\small
\setlength{\tabcolsep}{4pt}
\renewcommand{\arraystretch}{1.1}
\begin{tabular}{lccccc}
\hline
Configuration & WTR & NRIE & WBMC & WBOD & WNUC \\
\hline
LLM-only      & 99.64 & 76.96 & 68.14 & 83.59 & 70.17 \\
Vision-only   & 99.79 & 72.84 & 73.76 & 84.66 & 68.18 \\
Full-FT       & 99.88 & 74.13 & 75.15 & 89.73 & 72.47 \\
Grayscale     & 89.49 & 68.73 & 51.63 & 79.43 & 49.45 \\
25\,\% data   & 99.28 & 64.63 & 57.42 & 71.27 & 55.03 \\
50\,\% data   & 99.71 & 68.93 & 62.85 & 77.69 & 63.60 \\
Naive uniform & 38.74 & 27.96 & 51.39 & 48.42 & 30.67 \\
\hline
\end{tabular}
\end{table}

\begin{figure}[t!]
\centering
\includegraphics[width=0.8\linewidth]{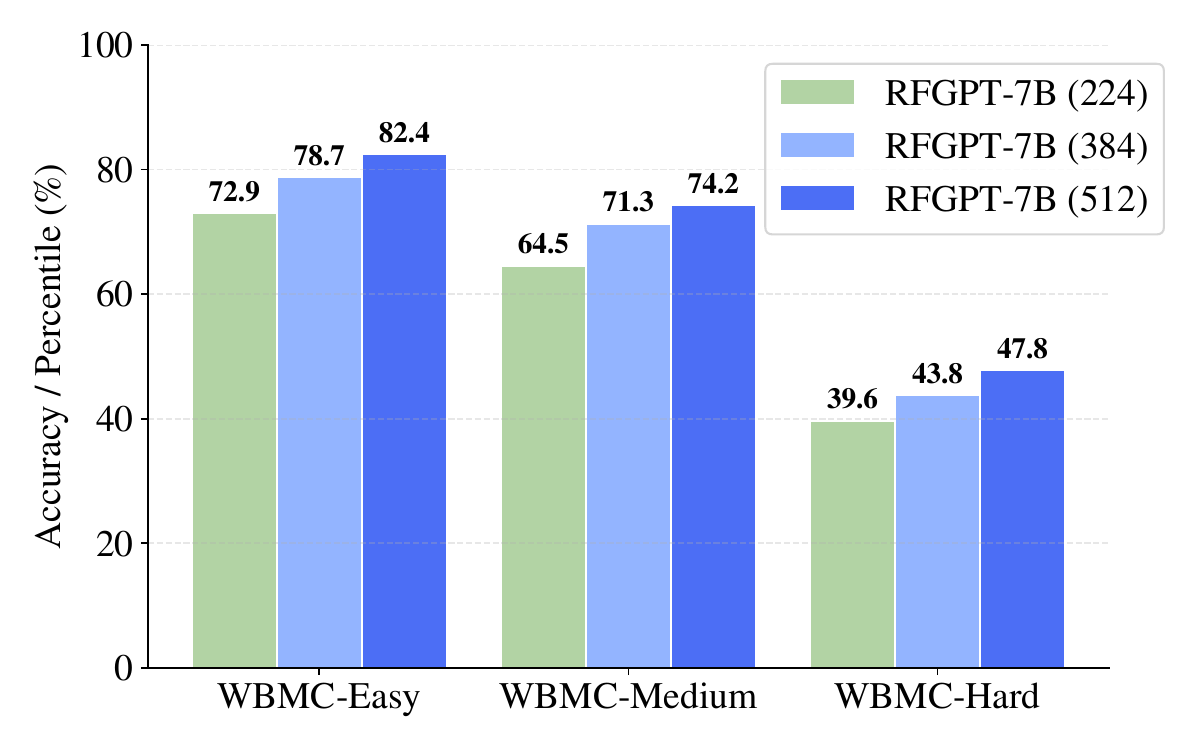}
\vspace{-3mm}
\caption{WBMC accuracy of RF-GPT-7B under different image resolutions.}
\label{fig:WBMC_image_resolution_ablation}
\end{figure}

\textbf{Image resolution.} We analyze the impact of input image resolution on WBMC. We evaluate \textbf{RF-GPT-7B} at \(224\), \(384\), and \(512\), as shown in Fig.~\ref{fig:WBMC_image_resolution_ablation}. The performance improves consistently as resolution increases, where WBMC-Easy rises from \(72.94\%\) through \(78.72\%\) to \(82.41\%\), WBMC-Medium from \(64.52\%\) through \(71.26\%\) to \(74.24\%\), and WBMC-Hard from \(39.64\%\) through \(43.76\%\) to \(47.77\%\). From \(224\) to \(512\), the absolute gains are \(+9.47\), \(+9.72\), and \(+8.13\) points on Easy, Medium, and Hard, respectively. These results indicate that higher-resolution spectrograms preserve more discriminative time--frequency details and benefit both coarse and fine-grained modulation recognition. However, increasing image resolution also increases RF token count, leading to higher memory/computation cost and longer inference latency.

\begingroup
\textbf{Synthetic-to-real validation.} A parallel follow-up study~\cite{bara2026rfanalyzer} evaluates an \emph{unmodified} RF-GPT on live SDR captures, five in-distribution OTA scenarios and one OOD shielded low-SNR Bluetooth burst, against Qwen2.5-VL and Llama 3.2 Vision in Table~\ref{tab:sim_to_real}. It reports two metrics: the Physical Attribute Extraction Score (PAES), the fraction of correctly extracted RF attributes; and the prompt-leakage rate, the fraction of bandwidth answers that parrot the prompt's sample-rate hint rather than measuring it from the spectrogram. RF-GPT attains $96\,\%$ PAES in-distribution versus $32\,\%$/$60\,\%$, retains $40\,\%$ on the OOD burst where both general models collapse to $0\,\%$, and shows the lowest prompt-leakage rate at $15\,\%$ versus $55\,\%$/$70\,\%$. The synthetic-trained spectrogram priors therefore transfer to real captures. \textcolor{black}{These numbers are reported by that study and not by a measurement campaign conducted for this paper. It captures each of the six scenarios three times, eighteen captures in total, using an Ettus USRP B210 with a broadband log-periodic antenna, at carriers from $98$ to $2437$\,MHz and sample rates from $5$ to $40$\,MHz; the OOD case is a Bluetooth mouse inside a commercial RF-shielding bag. PAES scores five binary attributes per trial, namely temporal class, spectral occupancy, SNR class, signal isolation and technology family, which we report normalized to a percentage.}
\endgroup

\begin{table}[t!]
\centering
\caption{Sim-to-real transfer on live SDR captures~\cite{bara2026rfanalyzer}.}
\label{tab:sim_to_real}
\begingroup
\small
\setlength{\tabcolsep}{4pt}
\renewcommand{\arraystretch}{1}
\centering
\resizebox{\columnwidth}{!}{%
\begin{tabular}{lccc}
\hline
Metric                         & RF-GPT      & Qwen2.5-VL & Llama 3.2-V \\
\hline
PAES, in-dist.\ OTA            & \textbf{96} & 32         & 60          \\
PAES, OOD low-SNR (shielded)   & \textbf{40} &  0         &  0          \\
Prompt-leakage rate            & \textbf{15} & 55         & 70          \\
\hline
\end{tabular}}
\endgroup
\end{table}

\textbf{Unseen-Task Benchmark.} To test whether RF-GPT generalizes beyond the predefined training tasks, we introduce a small \emph{Unseen-Task Benchmark} (UTB) of three question types verifiably absent from the training instructions of Sec.~\ref{sec:rf_grounding}, with ground truth derived deterministically from existing scene metadata: Q1 binary coexistence detection, Q2 integer modulation-diversity count, and Q3 binary scene-level burst pattern. As reported in Table~\ref{tab:utb}, RF-GPT cleanly separates from the general VLMs on two of the three: RF-GPT-7B reaches $100\%$ on Q3 against the $50\%$ random-guess floor, and RF-GPT-3B/-7B reach $98.8\%$ within-one on Q2 versus $56.7\%$ for the general VLM. \textcolor{black}{Q2 is scored within-one of the true count rather than by strict exact match.} Q1 sits nearer the floor, yet RF-GPT-7B still reaches $31\%$ YES-class recall versus $5\%$ for the base model. The trained RF-GPT models outperform the general VLMs on all three. \textcolor{black}{All three question types are absent from the training instructions and none requires a task-specific head, so the UTB shows that RF-GPT's instruction-following interface extends beyond the trained task set. With three question types and a modest Q1, we present it as a first probe of that behavior rather than an exhaustive characterization.}

\begin{table}[t!]
\centering
\caption{Unseen-Task Benchmark accuracy (\%).}
\label{tab:utb}
\small
\setlength{\tabcolsep}{4pt}
\renewcommand{\arraystretch}{1}
\resizebox{\columnwidth}{!}{%
\begin{tabular}{lccc}
\hline
UTB question                    & RF-GPT-7B & RF-GPT-3B & Qwen2.5-VL-7B \\
\hline
Q1 Coexistence detection        & 62.8      & 49.6      & 48.7          \\
Q2 Mod.-diversity count \textcolor{black}{(within one)} & 98.8      & 98.8      & 56.7          \\
Q3 Burst pattern (scene)        & 100.0     & 58.0      & 50.0          \\
\hline
\end{tabular}}
\end{table}

\textcolor{black}{\textbf{Phase-aware representation ablation.} To isolate the impact of phase information on a phase-sensitive task without changing the vision encoder, we hold the $[3,H,W]$ input interface fixed across all encoding variants. We compare the default \emph{Magnitude} encoding, a viridis colormap of the log-magnitude spectrogram, against two phase-aware encodings of the same complex STFT: \emph{Magnitude-Phase (Coherent)}, whose three channels are the dB-magnitude and the quadrature components of the unit-modulus phasor $(\cos\phi+1)/2$ and $(\sin\phi+1)/2$; and \emph{Magnitude-Phase (Polar)}, where magnitude (viridis) and phase (colorized) are placed side-by-side as two panels of the same image. RF-GPT-3B is fine-tuned on the WBMC training union under matched hyperparameters across the three configurations (Table~\ref{tab:phase_ablation}). Magnitude outperforms both phase-aware encodings, and Magnitude-Phase (Polar) substantially outperforms Magnitude-Phase (Coherent): Polar preserves the familiar viridis-magnitude as one of its two panels, whereas Coherent replaces all three channels with a mixed magnitude $+$ $(\cos\phi, \sin\phi)$ encoding that the pretrained vision encoder has not seen. A phase-aware front end therefore requires a custom-pretrained RF encoder, which in turn requires large-scale pretraining to close the gap. We leave the phase-aware extension as future work.}

\begin{table}[t!]
\centering
\captionsetup{font={color=black},labelfont={color=black}}
\caption{Magnitude vs phase-aware encodings, WBMC signal accuracy (\%); $\Delta$ in parentheses is vs Magnitude. RF-GPT-3B fine-tuned for 1 epoch under identical hyperparameters; only the per-pixel encoding differs across the three configurations.}
\label{tab:phase_ablation}
\small
\setlength{\tabcolsep}{4pt}
\renewcommand{\arraystretch}{1}
\resizebox{\columnwidth}{!}{%
\begin{tabular}{lccc}
\hline
WBMC split    & Magnitude    & Magnitude-Phase (Coherent)        & Magnitude-Phase (Polar)         \\
\hline
WBMC-Easy     & 63.04  & 26.38 \,($-36.66$)         & 52.49 \,($-10.55$)        \\
WBMC-Medium   & 47.43  & 13.82 \,($-33.61$)         & 36.21 \,($-11.22$)        \\
WBMC-Hard     & 22.54  &  5.52 \,($-17.02$)         & 16.26 \,($-6.28$)         \\
\hline
\end{tabular}}
\end{table}

\section{Conclusion and Future Work}
\label{sec:conclusion}

In this paper, we introduce the concept of RFLM and present RF-GPT as its first realization. RF-GPT integrates RF spectrograms into a multimodal LLM, enabling a unified interface between RF signals and natural language. Through adopting realistic synthetic waveform generation, deterministic spectrogram captioning, and LLM-based instruction synthesis, we constructed a large RF instruction-tuning dataset spanning multiple technologies and tasks. On our comprehensive benchmarks, RF-GPT substantially outperformed general-purpose VLMs that exhibit very limited RF-specific understanding under our benchmark setting, and showed competitive performance with strong CNN and Transformer baselines, while additionally offering natural-language answers, structured outputs, and explanations from a single model.

Future directions include extending RF-GPT with phase-aware and multi-antenna front ends, such as complex-spectrogram, beam-space, or CSI-grounded tensors, to broaden the task families it can serve, building an over-the-air counterpart of the present benchmark suite for in-the-wild validation, and integrating RF-GPT into spectrum-monitoring and AI-native network management prototypes, moving RFLMs from synthetic proof-of-concept toward practical RF intelligence engines for next-generation wireless networks.

\appendices

\begingroup\color{black}
\section{Implementation Details}
\label{app:impl}

\textbf{Generator parameter ranges.} We do not use ad hoc RF parameter ranges. For each technology, we start from the MATLAB toolbox example/default configurations and randomize documented, standard-constrained fields exposed by the corresponding generator APIs. Every accepted waveform must pass the toolbox validation logic for the underlying standard: 3GPP for 5G NR, LTE, and UMTS; IEEE 802.11ax/be for WLAN; ETSI EN 302 307 for DVB-S2; IEEE 802.15 for Bluetooth. The exhaustive per-field option sets are therefore MATLAB/standard reference tables rather than independent modeling choices, so we do not reproduce those long API-level enumerations here. Instead, we report the experiment-level ranges that define the benchmark distribution: per-scene UE counts range from 1 for Bluetooth and DVB-S2 up to 5 for 5G NR, with SNR drawn from $[-5, 30]$\,dB. TorchSig wideband contributes a 57-class modulation taxonomy spanning AM, FM, ASK, FSK, OOK, OFDM, PSK, QAM, tone and chirp families, with 2--5 co-existing signals per scene. \textbf{Splits/seeds.} Approximately 5000/100/1000 train/val/test scenes per toolbox technology label, where NR (DL, UL) and WLAN (AX, BE) each contribute two labels; 40000/1000/2000 each for two TorchSig wideband variants,  where WBMC uses the no-overlap variant; WBOD uses the with-overlap variant. Deterministic per-technology seeding ensures train, validation, and test scenes are disjoint and reproducible.

\textbf{Prompt templates.} One example per benchmark family is reproduced below; the complete set will be released together with the evaluation benchmarks.
\begin{itemize}
\item \textbf{WBMC-Medium:} \emph{``Your task is to analyze an RF spectrogram and identify all modulation classes present in the scene. For this particular example, the only possible modulation classes are: [per-scene allowed list]. Output one modulation label per signal as a JSON array of strings, with no extra text.''}
\item \textbf{WBOD-Easy:} \emph{``Look at the wideband RF spectrogram and decide if any pair of signals overlaps in time, frequency, or both. Answer with exactly one of: `neither', `time-only', `frequency-only', `both'. Return only the label string, with no extra text.''}
\item \textbf{WTR-Easy:} \emph{``Your task is to analyze an RF visualization and determine the most likely wireless technology from a given list. Here are the possible classes: [`bluetooth', `dvbs2', `lte', `nr', `umts', `wlan']. Your response must contain the exact name of the class only.''}
\item \textbf{WNUC-AX-Medium:} \emph{``Your task is to examine the RF visualization and choose the user-count bucket that best matches the scene. Each bucket spans 10 users. Select one bucket from: [`1-10', `11-20', \ldots, `91-100']. Respond with the bucket string exactly.''}
\item \textbf{NRIE (SCS attribute):} \emph{``Your task is to determine the subcarrier spacing (SCS) in kHz of the 5G NR carrier shown in this spectrogram. Possible SCS values (in kHz) are: [`15', `30', `60', `120', `480', `960']. Respond with exactly one of these numbers as a string.''}
\end{itemize}

\textbf{Parsing.} Per-task: relaxed string match for WTR, NRIE categorical and WBOD-Easy; Markdown-fence-tolerant JSON-array extraction for WBMC and WBOD-Medium; integer extraction with MAE for WNUC-Hard and NRIE counting.

\textbf{Release.} We plan to release the evaluation benchmarks, namely the instruction sets, scorers, parsing rules and prompt templates, at the appropriate time. Generator code, training pipeline, and instruction-synthesis code are not part of the public release.
\endgroup

{\scriptsize
\bibliographystyle{IEEEtran}
\bibliography{ref}}

\end{document}